\documentclass[sigconf,nonacm]{acmart}
\usepackage[utf8]{inputenc}

\usepackage[utf8]{inputenc}
\usepackage{graphicx} % Required for inserting images
\usepackage{comment}
\usepackage{tikz}
\usepackage{graphicx}
\usepackage{xspace}
\usepackage{xcolor}
\usepackage{xstring}
\usepackage{color, colortbl}
\usepackage{array}
\usepackage{multicol,multirow}
\usepackage{enumitem}
\usepackage{booktabs}
\usepackage{caption}
\usepackage{subcaption}
\usepackage{fontawesome}
\usepackage{cleveref}
\usepackage{balance}
\usepackage{hyperref}
\usepackage{soul}
\usepackage{xurl}
\usepackage{natbib}
\usepackage{listings}
\usepackage{balance}
\usepackage{makecell}
\usepackage{amsfonts}
\usepackage{amsmath}
\usepackage{microtype}
\usepackage{nicematrix}
\usepackage{relsize}

\usepackage[framemethod=tikz]{mdframed}

\usepackage{longtable}

\usepackage{wasysym}

\definecolor{darkpastelgreen}{rgb}{0.01, 0.75, 0.24}
\definecolor{cadmiumgreen}{rgb}{0.0, 0.42, 0.24}
\definecolor{brickred}{rgb}{0.8, 0.25, 0.33}
\definecolor{cornellred}{rgb}{0.7, 0.11, 0.11}
\definecolor{burgundy}{rgb}{0.5, 0.0, 0.13}
\definecolor{frenchblue}{rgb}{0.0, 0.45, 0.73}
\definecolor{light-gray}{gray}{0.92}
\definecolor{lightlight-gray}{gray}{0.97}
\definecolor{codegray}{gray}{0.90}
\definecolor{inputgray}{gray}{0.90}

\definecolor{dkgreen}{rgb}{0,0.6,0}
\definecolor{light-green}{rgb}{0,0.5,0}
\definecolor{applegreen}{rgb}{0.55, 0.71, 0.0}
\definecolor{lightgreen}{rgb}{0.56, 0.93, 0.56}
\definecolor{gray}{rgb}{0.5,0.5,0.5}
\definecolor{mauve}{rgb}{0.58,0,0.82}
\definecolor{apricot}{rgb}{0.98, 0.81, 0.69}
\definecolor{bubblegum}{rgb}{0.99, 0.76, 0.8}

\global\mdfdefinestyle{rtboxstyle}{%
linecolor=black,%
leftmargin=0cm,rightmargin=0cm,linewidth=0.4pt,
roundcorner=2, skipabove=0.5em, innerleftmargin=5pt, innerrightmargin=5pt,
skipbelow=0pt,backgroundcolor=lightlight-gray
}

\newcommand{\rtbox}[1]{\begin{mdframed}[style=rtboxstyle]{{#1}}\end{mdframed}}

\newcommand{\cc}[1]{\mbox{\smaller[0.5]\texttt{#1}}}

\newcommand{\ie}{\textit{i}.\textit{e}.}
\newcommand{\eg}{\textit{e}.\textit{g}.}

\if 0

\fi

\newif\ifdraft\drafttrue
\newif\ifnotes\notestrue
\ifdraft\else\notesfalse\fi

\input{glyphtounicode}
\newcommand{\squishlist}{
\begin{itemize}[noitemsep,nolistsep]
  \setlength{\itemsep}{-0pt}
}
\newcommand{\squishend}{
  \end{itemize}
}

\setlist[enumerate]{leftmargin=.25in, topsep={2pt}, partopsep={0pt}}

\newcommand{\PP}[1]{
\vspace{2px}
\noindent{\bf \IfEndWith{#1}{.}{#1}{#1.}}
}

\newcommand{\PPNS}[1]{
\noindent{\bf \IfEndWith{#1}{.}{#1}{#1.}}
}

\newcommand{\boxbeg}{
\vspace{2px}
\noindent\begin{tabular}{|l|}\hline
\begin{minipage}{3.2in}
\vspace{2px}
\noindent
}

\newcommand{\boxend}{
\vspace{2px}
\end{minipage}\\ \hline
\end{tabular}
\vspace{-10pt}
}

\everypar{\looseness=-1}

\AtBeginDocument{%
  \providecommand\BibTeX{{%
    \normalfont B\kern-0.5em{\scshape i\kern-0.25em b}\kern-0.8em\TeX}}}

\setcopyright{none}
\renewcommand\footnotetextcopyrightpermission[1]{} % removes footnote with conference information in first column
\begin{document}
\sloppy

% \title{DNS Abuse for Phishing}
% \title{Anatomy of DNS Abuse in Phishing: From Registration to Detection, Brand Impersonation, and DNS Insights}
% \title{From Inception to Exposure: Navigating DNS Abuse, Brand Squatting, and Phishing Lifecycles}
% \title{Tracing the Shadows of DNS Abuse: Brand Impersonation, Domain Lifespans, and Phishing Tactics}
% \title{From Registration to Detection: Unveiling DNS Abuse in Phishing}
% \title{Unveiling DNS Abuse in Phishing: Lifespans of Maliciously-Registered Phishing Domains}
% \title{From Registration to Detection: Lifespans of Maliciously-Registered Phishing Domains}
% \title{DNS Abuse in Phishing: Malicious Registrations and Lifespans to Detection}
% \title{Beyond Blocklists: Unveiling DNS Abuse in Phishing Domains}
\title{Registration, Detection, and Deregistration: Analyzing DNS Abuse for Phishing Attacks}

\author{Kyungchan Lim}
\affiliation{%
  \institution{University of Tennessee}
  \city{Knoxville}
  \country{USA}}
\email{klim7@utk.edu}

\author{Raffaele Sommese}
\affiliation{%
  \institution{University of Twente}
  \city{Enschede}
  \country{The Netherlands}}
\email{r.sommese@utwente.nl}

\author{Mattijs Jonker}
\affiliation{%
  \institution{University of Twente}
  \city{Enschede}
  \country{The Netherlands}}
\email{m.jonker@utwente.nl}

\author{Ricky Mok}
\affiliation{%
  \institution{CAIDA/UC San Diego}
  \city{San Diego}
  \country{USA}}
\email{cskpmok@caida.org}

\author{kc claffy}
\affiliation{%
  \institution{CAIDA/UC San Diego}
  \city{San Diego}
  \country{USA}}
\email{kc@caida.org}

\author{Doowon Kim}
\affiliation{%
  \institution{University of Tennessee}
  \city{Knoxville}
  \country{USA}}
\email{doowon@utk.edu}

\renewcommand{\shortauthors}{Kyungchan Lim et al.}

\begin{abstract}
Phishing continues to pose a significant cybersecurity threat.
While blocklists currently serve as a primary defense, due to their reactive, passive nature,  these delayed responses leave phishing websites operational long enough to harm potential victims.
It is essential to address this fundamental challenge at the root, particularly in phishing domains.
Domain registration presents a crucial intervention point, as domains serve as the primary gateway between users and websites.

We conduct a comprehensive longitudinal analysis of 690,502 unique phishing domains, spanning a 39-month period, to examine their characteristics and behavioral patterns throughout their lifecycle—from initial registration to detection and eventual deregistration. 
% We analyze 690,502 unique phishing domains collected over a 39-month period. 
We find that 66.1\% of the domains in our dataset are maliciously registered, leveraging cost-effective TLDs and targeting brands by mimicking their domain names under alternative TLDs (\eg, \cc{.top} and \cc{.tk}) instead of the TLDs under which the brand domains are registered (\eg, \cc{.com} and \cc{.ru}). 
We also observe minimal improvements in detection speed for maliciously registered domains compared to compromised domains. 
Detection times vary widely across blocklists, and phishing domains remain accessible for an average of 11.5 days after detection, prolonging their potential impact. 
Our systematic investigation uncovers key patterns from registration through detection to deregistration, which could be leveraged to enhance anti-phishing active defenses at the DNS level.

\end{abstract}

\maketitle

\section{Introduction}
\label{sec:intro}

Phishing attacks continue to pose one of the most pervasive cybersecurity threats, with attackers deploying increasingly sophisticated impersonation tactics. 
The attackers create convincing replicas of legitimate websites (\eg, \cc{facebook}.\cc{com} or \cc{USPS}.\cc{com}), to deceive users into divulging their login credentials and sensitive information.
Such attacks have substantial consequences, leading to financial losses for victims~\cite{IC3_Report_2024}, reputational harm for impersonated organizations~\cite{Reputational_damages}, and compromised business infrastructures~\cite{moura2024characterizing}.

% Phishing attacks remain a pervasive cyber threat, with a record number of 597,789 unique phishing in April 2023\DK{what specific number}\KL{Added highest number from report} of incidents reported in the second quarter of 2023~\cite{docsapwg72:online}.
% % ~\cite{docsapwg95:online}. 
% Phishing attackers generate deceptive websites (\eg, facebook.com or USPS.com) to lure victims into revealing their credentials.

% The significant rise in phishing attacks underscores the ongoing challenges posed by increasingly sophisticated phishing websites, phishing kits, and evasion techniques~\cite{add}. 

The blocklisting mechanisms (\eg, Google Safe Browsing~\cite{SafeBrow22:online}) currently serve as the primary defense against phishing attacks.
Google Safe Browsing is integrated into Google Chrome browsers and by default enabled for end-users.
These systems aim to protect users by preventing access to known (\ie, blocklisted) phishing websites. 
However, their reactive (\ie, passive) nature introduces critical security gaps in phishing protection.
The fundamental limitation of blocklists lies in their update latency---the time gap between when attackers register domains and deploy a new phishing site and when security crawlers detect, verify, and add it to the blocklist.
This delay creates a vulnerability window during which new phishing sites remain accessible to potential victims, allowing attackers to freely operate their campaigns.
Notably, a previous work~\cite{oest2020sunrise} indicated that 75\% of victims may encounter the malicious site before blocklist updates take effect.

% To effectively combat phishing attacks, it is essential to address the fundamental problem at the root---DNS (Domain Name System) level, particularly phishing domains.
To effectively combat phishing attacks, it is essential to address the fundamental problem at the root, particularly phishing domains.
Domains play a pivotal role in connecting users to websites, including malicious ones.
This critical position makes domains an ideal intervention point for detecting and preventing phishing attacks before they can reach potential victims.
% \KL{Add importance of analyzing Maliciously registered domain}
% Unfortunately, prior work
% \DK{prior work summary and limitation.}
% 
Particularly, phishing attackers can choose between two strategies for utilizing domain names: 1) registering a new domain specifically for malicious purposes; or 2) compromising an existing, legitimate website with an already established domain. 
Maliciously registered domains present a unique opportunity for mitigation at the domain level, as these domains are intentionally created to facilitate malicious activities. 

% \KL{summarize COMAR, ccTLD papers, start with specifically,... what previous papers looked, what has been missing }
% \KL{this area is not studied before even though it is important area...}
Prior studies~\cite{maroofi2020comar,moura2024characterizing,hao2013understanding,oest2019phishfarm,oest2020phishtime,oest2020sunrise} have explored various aspects of phishing websites, such as ccTLD, URL patterns and visual content.
Specifically, Moura et al.~\cite{moura2024characterizing} analyzed phishing domains mimicking target brand webpages but focused solely on three European ccTLDs: \cc{.nl}, \cc{.ie}, and \cc{.be}.
While Maroofi et al.~\cite{maroofi2020comar} introduced the methods to define maliciously registered domains,
the characteristics of maliciously registered domains for phishing attacks have been unexplored.
% the scope for expanding their approach remains limited 
% \DK{what limited?}\KL{updated, application? implication?}
Despite the importance of understanding the dynamic behaviors and lifecycle of maliciously registered domains, these aspects remain largely unexamined to date.\looseness=-1
% 
% Understanding how domains are registered and utilized by attackers is critical for mitigating phishing attacks. %Phishing attacks continue to be a significant cybersecurity threat, with attackers employing sophisticated impersonation tactics to deceive users.
% While prior studies~\cite{maroofi2020comar,hao2013understanding} have explored this criterion, the prevalence of maliciously registered domains continues to grow compared to compromised domains~\cite{Phishing18:online}. 
% Effectively combating phishing attackers and mitigating the registration of malicious domains requires a systematic understanding of how phishing domains are registered and the underlying strategies attackers employ.
%
% registering a malicious domain or compromising an existing website with an already established domain.
% A maliciously registered domain can be mitigated in domain level since such domains are registered for a malicious purpose.
% Previous works~\cite{add} have looked into this criteria however malicious domains still grows when compare to compromised domains~\cite{Phishing18:online}.
% To combat phishing attackers and mitigate registering malicious domains, we need a systematic understanding of how phishing attackers are registering domains maliciously.
%
% \KL{need to be more specific}

To address this gap, our study undertakes a systematic, longitudinal analysis of phishing attacks using phishing domains, with a focus on those that are maliciously registered domains. 
By examining these phishing domains from registration to detection and eventual deregistration, we aim to better understand the phishing attack ecosystem at the domain level. 
% Specifically, we identify maliciously registered phishing domains and analyze their distinctive characteristics, such as the abuse of specific TLDs, targeted brands, and configurations of DNS records. 
% Additionally, we investigate the lifespan of these maliciously registered domains, tracking their trajectory from registration to detection, and beyond, until deregistration.
% 
To further understand maliciously registered domains, we raise the two following research questions.
% \KL{remove rq1 and keep only rq2 and rq3}
% \textbf{RQ1:} \textit{How can we effectively identify maliciously registered domains?},
% , and what insights can be drawn from their characteristics and behaviors?
\textbf{RQ1:} \textit{What are the characteristics of maliciously registered domains and how can we find maliciously registered domains?}
\textbf{RQ2:} \textit{What is the lifecycle of a maliciously registered domain?}
Our analysis shows that 66.1\% of all names in our phishing domains dataset are specifically registered for malicious purposes.  
To better understand these malicious domains, we examine their characteristics, focusing on TLD usage and targeted brands. 
We observe that new gTLDs (\eg, \cc{.top}, \cc{.shop}) are widely utilized due to their low cost (as little as \$1 per domain). 
Following the cessation of Freenom, the use of the \cc{.cn} TLD increased significantly. 
Notably, the \cc{USPS} brand experienced a sharp rise in domain registrations, frequently under cost-effective TLDs. The latter holds true for \cc{Ozon} as well.
Our observations align with prior research~\cite{Phishing18:online}.

To gain deeper insights into phishing domains with malicious registration activity, we analyze their DNS records, dynamic behavior, and lifespan, spanning from registration to detection and eventual deregistration. Our analysis reveals that phishing domains often exhibit dynamic DNS behavior, frequently updating their DNS records with short TTLs, indicative of fast-flux DNS techniques. Regarding lifespan, maliciously registered domains are detected slightly faster than compromised domains with a median detection time of 16.3 days for malicious domains compared to 86 days for compromised domains.
% \MJ{16.3 to 86 is considered ``marginal``??} \KL{removed marginal} 
On average, deregistration occurs approximately 11.5 days after detection. However, detection delays vary significantly across blocklists, with some domains listed in \cc{Phishing.Database} showing an average detection delay of up to 388.5 days.

Our contributions are as follows:\looseness=-1
\begin{itemize}[leftmargin=*, topsep=0pt, itemsep=0em]
% \KL{from our dataset, compared to other dataset, similar to other(previous) report}
\item Building on previous methods, we enhance the approach to identify maliciously registered domains. Our analysis reveals that 66.1\% of the domains in our dataset are maliciously registered, with the remainder being compromised domains.
\looseness=-1
% Other than maliciously registered domains defined by COMAR~\cite{maroofi2020comar} method, we also find that 28.2\% domains are creating phishing domains with randomly generated characters. 

\item From our analysis of maliciously registered domains in our dataset, we identify three key characteristics:
1) New gTLDs: Domains frequently use low-cost new gTLDs (\eg, \cc{.top}, \cc{.xyz}, and \cc{.online}), with \cc{.cn} usage rising after Freenom ceased free registrations~\cite{SuedbyMe25:online}, aligning with previous reports~\cite{Phishing18:online}.
% Use of new gTLDs: new gTLDs (\eg, \cc{.top}, \cc{.xyz}, \cc{.shop}, and \cc{.online}) are frequently used in maliciously registered domains. 
% Additionally, after Freenom ceased offering free registrations~\cite{SuedbyMe25:online}, the use of \cc{.cn} in malicious domains increased significantly which aligns with previous reports~\cite{Phishing18:online}.
2) TLD Variation in Brand Targeting: Phishing domains targeting brands (\eg, \cc{USPS}, \cc{OZON}) often use alternative TLDs (\eg, \cc{.top}, \cc{.tk}) instead of the brand’s original TLDs (\eg, \cc{.com}, \cc{.ru}).
% 
% TLD Variation in Brand Targeting: Phishing domains impersonating popular brands (\eg, USPS and OZON) commonly use alternative TLDs (\eg, \cc{.top} and \cc{.tk}) instead of the original brand’s TLD (\eg, \cc{.com} and \cc{.ru}).
3) DNS Fast Flux: 64.3\% of domains show frequent DNS updates, with 25.8\% using TTLs below 3600 seconds to possibly enable DNS fast flux.
% 
% Rapid DNS Record Changes: 64.3\% of domains exhibit frequent DNS record changes, with 25.8\% of domains set TTL value less than 3600 (1 hour) which can be used for a tactic associated with DNS fast flux.
% \item We find that by different impersonated brands, detection time varies, for example, USPS and Ozon are detected within XX days while Microsoft is detected within XX days. Also, we find various characteristics of impersonating brands, for example, attackers targeting Ozon uses Freenom TLD (\eg, .tk) the most. 

\item We find that maliciously registered domains are detected by blocklists (\eg, APWG) faster than compromised domains, with a median detection time of 16.3 days for malicious domains compared to 86 days for compromised domains. 
Additionally, detection times vary significantly across blocklists, with USPS and Ozon being the quickest with 1.4 and 1.3 days respectively. 
Even after detection by blocklists (\eg, APWG), phishing domains remain accessible for an average of 11.5 days, prolonging the risk to potential victims.
% \item We find that maliciously registered domains stay alive longer even after it is detected (23 days compared to 11.5 days). From those of maliciously registered domains, an average lifetime is 286 days. The average detection time of maliciously-registered domains are not much shorter than compromised domains by XX days. \KL{All detection time - maliciously-registered detection time}
\item We present a comprehensive longitudinal analysis of phishing domains (39 months). We publicly share our collected phishing dataset (\ie, phishing domains) to facilitate future phishing research upon acceptance.
% \KL{direction to share data, or share domain}
\end{itemize}

\section{Background}
\label{sec:background}
We provide a brief overview of domain registration and DNS records, with an emphasis on phishing attacks.

% \begin{figure}[!t]
%     \centering
%     \includegraphics[width=1\linewidth]{fig/overview.pdf}
%     \caption{Overview of Our DNS Analysis on Phishing Domains.\KL{Change section as our current version of section names. Add number of filtered domain}\KH{How about change Section X to RQ X?}}
%     \label{fig:overview}
% \end{figure}

\subsection{Domain Registration and DNS Record}

\PP{Domain Registration}
Domain registration is the foundational process through which a unique domain name is acquired and associated with an individual or organization. 
This process involves selecting a domain name and choosing a top-level domain (TLD), such as \cc{.com}, \cc{.org}, or country-specific TLDs, such as \cc{.us} or \cc{.cn}. 
Once a domain is registered through a registrar, critical DNS records---such as A records, which link the domain to an IP address, and NS records, which designate authoritative name servers---are established to facilitate the Web services. 
The registry maintains the TLD's zone file, which includes delegation details for domains under that TLD. 
These zone files are updated in real-time or periodically by the registry as domain registrations and configurations change. 
Separately, organizations like ICANN collect published snapshots of these zone files at regular intervals (\eg, every 24 hours for gTLDs~\cite{HelpCent91:online}), though the exact frequency depends on the TLD administrator’s policies. 
It’s important to note that the frequency of published zone file snapshots is distinct from the registry’s internal updates to the zone file.

% Then, the registrar submits delegation information to the registry, who manages the corresponding TLD zone files. 
% These zone files are typically updated and collected at regular intervals, often every 12 hours, although the exact collection frequency may vary depending on the TLD administrator's policies.
% \MJ{I think you're conflating TLD zone updates with the update interval of published zone file snapshots here} 
% For example, ICANN collects gTLD zone files every 12 hours~\cite{HelpCent91:online}.

\PP{WHOIS}
Registration data is typically accessed through WHOIS or the Registration Data Access Protocol (RDAP). 
WHOIS has been the standard for retrieving domain registration information since the 1970s. 
However, due to its inconsistencies and limitations, RDAP was introduced in 2015 as its successor. RDAP improves upon WHOIS by offering structured, machine-readable registration data along with advanced features such as differentiated access, internationalization, and extensibility.
By examining the domain registration choices of phishing sites, including their TLD preferences, registrar selection, and DNS configuration, researchers can uncover patterns that may inform more effective, proactive detection methods against these evolving threats.
\looseness=-1

\PP{Top-level Domain (TLD)}
As described in prior work~\cite{AtLargeT9:online}, gTLDs can be categorized into legacy gTLDs and new gTLDs. New gTLDs refer to TLDs introduced as part of ICANN's expansion program in 2012.
Initially, there were only 8 gTLDs, and another 8 in 2004.
% (\cc{.com}, \cc{.edu}, \cc{.gov}, \cc{.mil}, \cc{.org}, \cc{.net}, \cc{.int}, \cc{.arpa}) 
% before the year 2000, followed by the addition of 7 more in 2000 (\cc{.aero}, \cc{.biz}, \cc{.coop}, \cc{.info}, \cc{.museum}, \cc{.name}, \cc{.pro}) 
% 
% (\cc{.asia}, \cc{.cat}, \cc{.jobs}, \cc{.mobi}, \cc{.tel}, \cc{.travel}, \cc{.xxx}). 
In 2012, ICANN launched the new gTLD program, which aimed to provide greater flexibility for registrants to create unique and innovative website names. This initiative also alleviated the overcrowding in the legacy gTLD market, offering more options for domain registration. Since the program's introduction, over a thousand new gTLDs have been delegated to the root zone, significantly expanding the domain name landscape.
\looseness=-1

\PP{DNS Record}
DNS records are fundamental components of the Domain Name System (DNS), serving as mappings that enable domain names to link to specific internet resources, such as IP addresses, email servers, and authoritative name servers. Each DNS record type provides unique information and functionality essential for domain operation. For instance, \cc{A} records (Address records) link a domain to an IPv4 address, directing users to the correct server when they access a website. \cc{NS} records (Name Server records) specify which servers are authoritative for a domain, manage DNS queries, and ensure accurate routing. 
% MX records (Mail Exchange records) designate mail servers responsible for receiving emails on behalf of a domain, a key aspect for domains involved in email-based phishing. 

DNS records may differ depending on the geographic or network location, known as the vantage points, from which the DNS query is made. This variation occurs because DNS configurations can be adapted to present different responses based on the requester’s location,
% a technique often used for legitimate purposes, 
such as content delivery optimization or load balancing.

% However, phishing attackers also exploit this capability to evade detection by selectively serving malicious content. By querying DNS records from multiple global vantage points, researchers can detect discrepancies in DNS responses that may signal evasive behavior.
% \DK{I am not sure whether phishing attackers exploit it. } \KL{removed}

% \PP{Challenges in Traditional Detection Methods}

\subsection{Phishing Attack and Tactic}
Phishing attacks are a type of advanced social engineering where cybercriminals deceive victims into divulging sensitive information. 
Phishing attackers craft fake websites that closely resemble legitimate ones (\eg, Facebook or PayPal), deceiving victims into entering their credentials. 
% \DK{phishing attacks starts with registering domains and deploying websites}
% As phishing techniques continuously evolve, a relentless cat-and-mouse struggle ensues between attackers, who invent new ways to bypass detection, and security systems that depend on databases of previously identified phishing sites to stay ahead.

\PP{Phishing Tactics for Domain Registration}
The choice of the registrar and TLD can significantly impact a domain's visibility, cost, and accessibility, with certain TLDs (\eg, \cc{.tk} or \cc{.xyz}) often being cheaper or subject to less stringent registration requirements. 
Phishing attackers frequently take advantage of this aspect of domain registration, choosing low-cost or lenient TLDs to host their malicious sites in large numbers while minimizing expenses~\cite{moura2024characterizing}.
Additionally, some registrars (\eg,  \cc{Alibaba} \cc{Cloud}~\cite{AlibabaC38:online}) have minimal verification protocols, making it easier for attackers to quickly register multiple domains in bulk under anonymous or fabricated identities~\cite{NewUserD82:online}.
This practice allows attackers to operate on a large scale, using each domain temporarily until it is flagged or blocked by detection systems, then transitioning to newly registered domains.
\looseness=-1

\PP{Phishing Tactics for Domain Name}
When conducting phishing attacks, attackers employ various domain registration strategies to deceive users. 
They commonly use typosquatting, registering domains with subtle misspellings, such as \cc{paypaI}.\cc{com} (using a capital \cc{I} instead of \cc{L}) or missing letters such as \cc{goole}.\cc{com}.
Another tactic involves creating domain variations by adding words or modifying the structure, resulting in domains like \cc{paypal}-\cc{secure}-\cc{login}.\cc{com} or \cc{login}-\cc{paypal}.\cc{net}.
Attackers also abuse different top-level domains (TLDs), using alternatives like \cc{.co} or country-specific codes instead of the legitimate \cc{.com}.
\vspace{-10px}
% A more sophisticated approach is the homograph attack, where they register domains using similar-looking Unicode characters or mixed character sets—for example, using a Greek omicron in 'gοοgle.com' or Cyrillic characters in 'paypal.com'.

\begin{figure*}[!t]
    \centering
    \includegraphics[width=.98\linewidth]{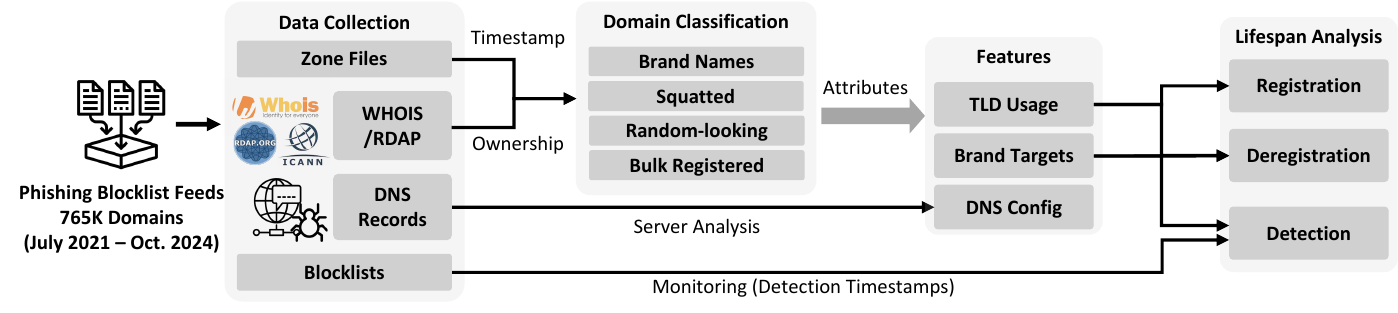}
    \vspace{-10px}
    \caption{Overview of Our DNS Analysis on Phishing Domains.}
    \label{fig:overview}
    % \vspace{-5px}
\end{figure*}

\section{Problem Statement}
\label{sec:motivation}

% \MJ{I think there's quite a bit of overlap/replication from the intro. I wonder if a dedicated problem statement section is altogether needed given that the problem statement, delta/novelty over related work, and research questions were all already presented to readers in the introduction.}

% Goal: How can we take a step further to detect phishing at the DNS level?
% , with attackers constantly evolving their tactics to evade detection and maximize victim reach. 
Phishing remains a major security threat, with traditional blocklist-based defenses (\eg, Google Safe Browsing---Google Chrome default anti-phishing system) suffering from significant detection delays. These systems often take hours or days to update after new phishing domains are registered, creating a critical window of vulnerability during which attackers can successfully target victims~\cite{lin2021phishpedia,liu2022inferring,liu2023knowledge,liu2024less,oest2019phishfarm,oest2020phishtime}.

Addressing phishing at the DNS level---when domains are first registered---is crucial, as domains are the primary gateway to phishing websites. However, while previous studies~\cite{maroofi2020comar,moura2024characterizing,hao2013understanding,oest2019phishfarm,oest2020phishtime,oest2020sunrise}  have focused on URL patterns, visual content, and blocklist data, there is limited understanding of how attackers exploit DNS registration strategies for phishing attacks.
To this end, our work aims to bridge this knowledge gap by focusing on maliciously registered domains for phishing attacks and their abuse of DNS systems. 
Particularly, we seek to answer our research questions through our analysis using our dataset of phishing domains: 
\textbf{RQ1:} \textit{What are the characteristics of maliciously registered domains, and how can we find maliciously registered domains?} and 
\textbf{RQ2:} \textit{What is the lifecycle of a maliciously registered domain?}.
\section{Dataset Collection}
\label{sec:datacollection}

\begin{table}[t]
\caption{Overview of Our Collected Dataset from July 2021 to October 2024 (39 months). We collect a total of 2.3M phishing URLs and 765K domains.}
\vspace{-10px}
\label{tab:dataset}
\resizebox{0.9\linewidth}{!}{ 
\begin{NiceTabular}{l r r r}
\toprule
\multicolumn{1}{c}{\textbf{Type}} & \multicolumn{1}{c}{\textbf{\# URLs}} & \multicolumn{1}{c}{\textbf{\# Domains}} & \multicolumn{1}{c}{\textbf{\# TLD}}\\
\midrule
APWG~\cite{APWGTheA83:online} & 2,184,835 & 697,237 & 1,203\\
phishunt.io~\cite{phishunt94:online} & 262,755 & 66,743 & 598\\
PhishStats~\cite{PhishSta46:online} & 221,331 & 57,299 & 541\\
OpenPhish~\cite{OpenPhis99:online} & 115,804 & 26,127 & 480\\
Malware-filter~\cite{curbengh9:online} & 76,465 & 24,300 & 470 \\
PhishTank~\cite{PhishTan96:online} & 5,579 & 1,695 & 195\\
Phishing.Database~\cite{mitchell17:online} & 393 & 236 & 51\\
\midrule
Total (Distinct) &2,294,267  & 765,910 & 1,258 \\
\bottomrule
\multicolumn{4}{l}{-- APWG: collected from Jul. 15, 2021 to Oct. 31, 2024.}\\
% \multicolumn{4}{l}{and Oct. 31, 2024 (1204 days).}\\
\multicolumn{4}{l}{-- Others: collected from May 31, 2024 to Oct. 31, 2024.}\\
% \multicolumn{4}{l}{and Oct. 31, 2024 (153 days).}
\end{NiceTabular}
}
\vspace{-10px}
\end{table}

To address our research questions, we collect a dataset comprising phishing URLs (\autoref{sec:collection:phishing:domain}), DNS records using a custom-built crawler (\autoref{sec:dns_record_collection}), and registration timestamps of phishing domains to analyze their characteristics and lifespans (\autoref{sec:domain_registration_collection}).
% define maliciously registered domains (\autoref{sec:def_mal_registered_domains}),
% \DK{data collection section}
% analyzing domain characteristics (\autoref{sec:mal-reg-char}),
% domain registration/deregistration histories (\autoref{sec:lifespan}).
% , and DNS records (\DK{section ref}). 

% We collect a dataset of phishing webisites

\subsection{Phishing URL and Domain Collection}
\label{sec:collection:phishing:domain}
As shown in~\autoref{tab:dataset}, we first collect 2.3M phishing URLs and their associated 765K distinct domains (1,258 TLDs) across a 39-month period spanning July 2021 to October 2024 
% To ensure comprehensive data collection, we aggregate phishing URLs 
from multiple prominent phishing blocklists including APWG (Anti-Phishing Working Group)~\cite{APWGTheA83:online}, Malware-filter~\cite{curbengh9:online}, 
OpenPhish~\cite{OpenPhis99:online}, 
Phishing-Database~\cite{mitchell17:online}, 
phishunt.io~\cite{phishunt94:online},
PhishStats~\cite{PhishSta46:online},
and PhishTank~\cite{PhishTan96:online}.
% Other blocklists have unique URLs of 115,805 and unique domains of 84,600 with 664 unique TLDs.
These sources have been used to better understand the phishing ecosystem~\cite{oest2020phishtime,oest2019phishfarm,kim2021security,lim2024phishing,oest2020sunrise,8376206}.
Particularly, APWG is a global industry association of anti-phishing entities, including banks and financial services companies, Internet service providers, law enforcement agencies, and security vendors.
APWG maintains an extensive database of phishing URLs gathered from multiple sources.

\subsection{DNS Record Collection}
\label{sec:dns_record_collection}
% \DK{Summarize the purpose of this dataset collection in one sentence.}
To answer our research question 
(\textbf{RQ1:} \textit{What are the characteristics of maliciously registered domains, and how can we find maliciously registered domains?}),
% (\textbf{RQ2:} \textit{What are the characteristics of maliciously registered domains?}), 
we develop a comprehensive DNS data collection system to monitor and analyze how phishing attackers configure and modify their DNS settings across different geographic locations. 
Our system periodically collects DNS records types of our collected phishing domains 
(\ie, \cc{A}, \cc{AAAA}, \cc{NS}, \cc{MX}, \cc{TXT})
% (\eg, \cc{A}, \cc{AAAA}, \cc{NS}, \cc{MX}, \cc{TXT}, \cc{RRSIG}, \cc{DNSKEY}, \cc{DS}, \cc{NSEC}) 
to provide detailed insights into their behavior.
% \MJ{Why such a long "e.g." list? Which RRtype is not included in this list? I would either reduce the example to a few records, or go the "i.e." route and list all.}\KL{Updated}
% Our longitudinal data collection spans 39 months, capturing the complete DNS configuration lifecycle of phishing domains.

\PP{DNS Crawler Design}
\autoref{fig:overview} illustrates our data collection process. 
% \RM{Fig. 1 needs to have a word `crawler'}\KL{added crawler icon}
We implement a multi-threaded crawler designed for scalability and reliability, using concurrent processing to efficiently handle thousands of domain queries. The crawler maintains a connection pool for database access and implements file-locking mechanisms to prevent data corruption during parallel operations.
The crawler collects detailed DNS information using the \cc{dig} command with comprehensive parameters. 
This approach enables the recursive collection of DNS records, capturing all possible types. For reliability, our system implements a retry mechanism with exponential backoff, attempting each query up to 5 times before marking it as failed.

% \PP{Temporal Resolution}
Our crawler operates at 30-minute intervals, enabling us to capture both gradual changes and modifications in DNS configurations. This high-frequency polling is crucial for detecting dynamic DNS behaviors that phishing attackers might employ to evade detection, such as fast-flux DNS or rapid record updates.
The system stores DNS responses in a structured JSON format, organized by domain, timestamp, and vantage point.

Our data collection period spanned from June 6, 2024, to October 31, 2024. 
We gathered URLs from blocklist feeds and extracted their domains and subdomains to perform DNS queries during this period. 
Using our crawler, we collected a total of 94,798 domains, including subdomains, with 11,932 being unique domains.
% \RM{A bit confused here. How can DNS crawler collect URLs? i assume URLs are phishing URLs? but 4.1 said you collected 2.3M URLs. } \KL{clarified}

% An SQLite database tracks each domain-resolver pair's query status, retry attempts, and failure timestamps. This database helps maintain collection continuity and enables systematic analysis of DNS record evolution over time.
% This helps us to collect over 211 million DNS records.

\PP{Vantage Points}
Moreover, to detect location-based DNS configurations, we query DNS records from 10 geographically diverse DNS resolvers. These include global providers (Google, Cloudflare, Quad9, OpenDNS) and regional servers across six continents (Brazil, South Africa, UK, Australia, South Korea, US). This distributed approach reveals if phishing domains serve different DNS responses based on geographic location—a technique attackers might use to evade detection or target specific regions.
\vspace{-10px}
% To investigate this tactic, our crawler queries DNS records from multiple vantage points worldwide (\eg, Cloudflare, OpenDNS, and locations such as BR, ZA, UK, AU, and KR). 
% The vantage points used include IP addresses such as 143.110.169.182, 160.36.0.66, 170.64.147.31, 200.150.97.226, 208.67.222.222, 222.97.189.7, 41.23.216.150, 8.8.8.8, and 9.9.9.9. \KH{This sentence is too verbose, listing IPs. }
% This approach allows us to determine whether phishing attackers adapt their DNS settings based on the geographic location of potential victims, revealing evasion strategies that target specific vantage points.\KH{This seems unfounded. Please add more technical elements.}

\subsection{Domain Registration Collection}
\label{sec:domain_registration_collection}
To investigate the lifecycle of maliciously registered domains (\textbf{RQ2:} ``\textit{What is the lifecycle of a maliciously registered domain?}''), we collect registration information (including timestamps and registrars) of our collected phishing domains.
We first utilize WHOIS and the Registration Data Access Protocol (RDAP)~\cite{RDAPORG31:online} from registrars and registries as WHOIS and RDAP provides basic information, such as registrar names and domain registration/expiration dates.

% whois: 25,987 
% rdap: 436,176
% ctlog: 71,156
% domaintools: 27,929
% DNS Coffee: 526,867 (DZDB is inclusive in this data)

\PP{GDPR Restriction}
However, due to the European General Data Protection Regulation (GDPR), the registration timestamp and the registrant's information (such as their name, address, and phone number) can be unavailable to the public.
% From our collection of WHOIS, we have seen that domains can
To this end, we leverage the methodology of COMAR~\cite{maroofi2020comar} to obtain registration timestamps of the domains whose information is hidden.
% Specifically, by using CT logs, we can utilize the first timestamp of when the certificate was issued to a domain.
% Previous study~\cite{kim2021security} shows that TLS certificates are used for phishing attacks.\DK{<- does not make sense for the reason why we use CT for collecting timestamps.}
% We also leverage DNSDB, a passive DNS (pDNS) dataset provided by DomainTools~\cite{Introduc45:online}, to obtain the first-seen timestamps of domains in the pDNS collection.
Furthermore, we also utilize DNS Zone files from DNS Coffee~\cite{homeDNSC2:online} and DZDB~\cite{homeDZDB83:online} for more comprehensive registration timing analysis. 
These services daily collect and archive TLD Zone files from ICANN~\cite{AboutZon71:online}.
The Zone file data provides 
first-appearance and last-seen timestamps of domains; the last-seen timestamp indicates when a domain has been deregistered and is no longer active. 
We also utilize passive DNS (pDNS) data through DomainTools' Farsight DNSDB~\cite{Introduc45:online}. 
This dataset includes a first-seen timestamp, indicating the earliest recorded observation of a domain in the passive DNS.

% By comparing this timestamp with blocklist detection times, we determine whether a domain is deregistered before or after being flagged, providing critical insights into domain lifecycles and detection timelines.

\PP{Our Collected Registration Data}
Our dataset includes domain registration timestamps collected from various sources: WHOIS (25,987 domains), RDAP (436,176 domains), CT logs (71,156 domains), DomainTools (27,929 domains), and DNS Coffee (526,867 domains, inclusive of DZDB data).
In total, we have 526,954 registration timestamps for unique domains. 
% \RM{We have the registration timestamps of 526,954 unique domains?}\KL{yes}
% Specifically, we determine registration timestamps by analyzing multiple data sources: the Registration Data Access Protocol (RDAP)~\cite{RDAPORG31:online}, TLS certificates (CT Logs)\DK{cite}, and passive DNS first-seen timestamps~\cite{Introduc45:online}.
% Furthermore, we enhance this approach by 
% 
% More specifically, we begin by collecting timestamps from WHOIS and RDAP, both of which provide domain registration timestamps and associated information, such as the registrar. Additionally, we utilize zone files to extract both the first-appearance and last-seen timestamps of domains. The last seen timestamp indicates when a domain has been deregistered and is no longer active. By comparing this timestamp with blocklist detection times, we determine whether a domain was deregistered before or after being flagged, providing critical insights into domain lifecycles and detection timelines.
% 
In sum, our approach enables us to achieve 76.3\% (526,954 out of 690,502) coverage of our collected domains. While relying solely on registration timestamps from WHOIS and RDAP provides 62.4\% (431,011 domains) coverage, incorporating additional data sources such as zone files, pDNS data, and CT logs significantly improves the completeness of our timestamp data.

\vspace{-3px}
\section{Identifying Maliciously Registered Domains}
\label{sec:def_mal_registered_domains}
We first define maliciously-registered domains and then devise a method to identify the ones
%maliciously-registered domains 
for phishing attacks.
We further analyze the bulk registrations of phishing domains.
% Then, we conduct a comparative analysis of maliciously-registered domains and compromised domains.

% \begin{table}[t]
% \caption{Maliciously-registered Domain. Each steps are taken after removing the Tranco 1M~\cite{Aresearc32:online} list of domains (total of 689,492).}
% \label{tab:maliciously-registered-domain-categories}
% \vspace{-5px}
% \resizebox{0.9\linewidth}{!}{ 
% \begin{NiceTabular}{l r r}
% \toprule
% \multicolumn{1}{c}{\textbf{Type}} & \multicolumn{1}{c}{\textbf{\# of URLs}} & \multicolumn{1}{c}{\textbf{\# of Domains$^*$}} \\
% \midrule
% (1) Brand Name in Domain & 709,694 & 247,699 (35.9\%) \\
% (2) Squatted Domain & 472,320 & 180,468 (26.2\%) \\
% (3) Random-looking Domain & 283,366 & 194,099 (28.2\%) \\
% (4) Bulk-registered Domain & 69,599 & 54,787 \xspace\xspace(7.9\%) \\
% \midrule
% Mal. Total${^\dagger}$ & 1,406,525 & 455,525 (66.1\%) \\
% % \midrule
% % Total & 2,178,088 & 689,492 \xspace(100\%) \\
% \midrule
% \multicolumn{3}{l}{${^\ast}$: Due to the overlap, total domains are over 100\%}
% % \\
% % \multicolumn{3}{l}{${^\dagger}$: Total \# of Maliciously-registered domain.}
% \end{NiceTabular}
% }
% \vspace{-10px}
% \end{table}
% \KL{redo the experiment with all blocklists}

% \subsection{Identifying Maliciously-Registered Domains}
% \KL{Using COMAR method: Squatted domains (in domain, subdomain, path), Levenshtein distance}
% \KL{Maliciously-registered domain: Squatted domain, brand in domain.}
\PP{Def. of Maliciously-registered Domains}
Phishing domains can be classified into two categories: maliciously registered domains and compromised domains. 
A maliciously registered domain is intentionally purchased by an attacker for malicious purposes.
% , often involving brand impersonation or other deceptive tactics. 
In contrast, a compromised domain is a legitimate domain originally used for benign purposes, but attackers exploit vulnerabilities of web servers and inject malicious content (\eg, phishing pages) into the benign servers. 
Detecting maliciously registered domains at an early stage is a critical step in preventing phishing attacks effectively.

% To determine whether a domain is maliciously registered, previous works~\cite{maroofi2020comar,Phishing18:online,de2021compromised,canali2013role,hao2016predator} show different approaches to finding maliciously registered domains.
% One of the most promising and detailed classification methods is a method from COMAR~\cite{maroofi2020comar}.
% Specifically, we leverage lexical features from the method where lexical feature provides 95\% of accuracy from previous work~\cite{maroofi2020comar}.
% From COMAR method~\cite{maroofi2020comar}, the lexical feature includes 9 features. 

\PP{Identification of Maliciously-registered Domains}
We utilize Tranco 1M domains~\cite{Aresearc32:online} as a reference to filter out both legitimate domains and web hosting (or website builder) service domains from our collected phishing domains.
For example, while `\cc{blogspot}.\cc{com}' is a legitimate blogging service, attackers may create subdomains, such as `\cc{usps}-\cc{tracking}-\cc{service}.\cc{blogspot}.\cc{com}' for phishing purposes. 
%By using the Tranco list as a reference, we are able to remove these hosting platform-based phishing domains from our analysis of maliciously registered phishing domains.
After removing the platform-based phishing domains, our list remains 689,492 domains. 
% \RM{also mentioned the number of removed domains?}
\looseness=-1

% \KL{add Tranco removal, what we have removed}
% Tranco: 7,745
Furthermore, we leverage the previous approaches~\cite{maroofi2020comar,Phishing18:online,de2021compromised,canali2013role,hao2016predator} on finding maliciously registered domains.
Especially, COMAR~\cite{CompareP38:online} demonstrated a 95\% accuracy using lexical features and registration timestamps in domains.
The method from COMAR includes nine lexical features (\ie, presence of a brand name in the domain name, path part of URL, and misspelled target brand name in the domain name).
By merging those features and additional features we discovered, we design our method to detect maliciously registered domains by using the following four steps: (1) brand name in domains, (2) squatted domains, (3) random-looking algorithmic domains, and (4) bulk registered domains.
As shown in~\autoref{tab:maliciously-registered-domain-categories}, each steps are taken after removing the Tranco 1M~\cite{Aresearc32:online} domains from the total list of domains.
% Previous works~\cite{maroofi2020comar,Phishing18:online,de2021compromised,canali2013role,hao2016predator} have explored various approaches to identifying maliciously registered domains. 
% One of the most detailed and effective classification methods is COMAR~\cite{maroofi2020comar}, which demonstrates a promising 95\% accuracy using lexical features. 
% Leveraging this approach, we focused on the lexical feature set described in COMAR, which consists of nine distinct features.
% 
% Through heuristic observations, 
% Specifically, we streamline these lexical features into two primary categories: brand names embedded in domains and squatted domains. 
% These categories effectively capture common patterns used by attackers to craft malicious domains. 
% Additionally, we identified two other prominent strategies that extend beyond the lexical features in COMAR: random-looking algorithmically generated domains and bulk-registered domains. 

% However, from our heuristic observation, we combined features into two categories: squatted domains and brand names in domains.
% Aside from two categories previously found, we also find that random-looking algorithmic domain and bulk registered domains are also widely used method to register malicious domains.

\begin{table}[t]
\caption{Maliciously-registered Domain. Each steps are taken after removing the Tranco Top 1M~\cite{Aresearc32:online} list of domains (total of 689,492).}
\label{tab:maliciously-registered-domain-categories}
\vspace{-10px}
\resizebox{0.9\linewidth}{!}{ 
\begin{NiceTabular}{l r r}
\toprule
\multicolumn{1}{c}{\textbf{Type}} & \multicolumn{1}{c}{\textbf{\# of URLs}} & \multicolumn{1}{c}{\textbf{\# of Domains$^*$}} \\
\midrule
(1) Brand Name in Domain & 709,694 & 247,699 (35.9\%) \\
(2) Squatted Domain & 472,320 & 180,468 (26.2\%) \\
(3) Random-looking Domain & 283,366 & 194,099 (28.2\%) \\
(4) Bulk-registered Domain & 69,599 & 54,787 \xspace\xspace(7.9\%) \\
\midrule
Mal. Total${^\dagger}$ & 1,406,525 & 455,525 (66.1\%) \\
% \midrule
% Total & 2,178,088 & 689,492 \xspace(100\%) \\
\midrule
\multicolumn{3}{l}{${^\ast}$: Due to the overlap, total domains are over 100\%}
% \\
% \multicolumn{3}{l}{${^\dagger}$: Total \# of Maliciously-registered domain.}
\end{NiceTabular}
}
\vspace{-15px}
\end{table}

\PP{(1) Brand Name in Domain}
The first approach to identifying maliciously registered domains involves detecting brand names within the domain or subdomain (\eg, \cc{usps}-\cc{security}.\cc{example}.\cc{com}, or \cc{www}.\cc{usps}-\cc{security}-\cc{login}.\cc{com}).
To establish a comprehensive baseline, we curate a list of the top 1,000 most targeted brands in our collected datasets (\ie, APWG), covering 97\% of the domains in our dataset. Domains or subdomains containing any of these brand names are flagged as part of this category.
Our analysis reveals that 33.9\% of domains in the dataset incorporate brand names in their domain or subdomain, highlighting the prevalence of this tactic among phishing attackers. Detailed results for maliciously registered domains across all categories are summarized in~\autoref{tab:maliciously-registered-domain-categories}.

% The first category of finding maliciously registered domains is finding a brand name in a domain or subdomain.
% We start our list of brands with the most targeted 1,000 brands which covers 97\% of domains from our dataset.
% For this case, we find domains if a brand name is in the domain or subdomain.
% As a result, from our dataset, we find that 33.9\% of domains are using brand name in their domain or subdomain.
% The result of the maliciously registered domain is shown in~\autoref{tab:maliciously-registered-domain-categories}.

\PP{(2) Squatted Domain}
% Having squatted domain (brand) names in domain / subdomain (Top 200’s squatted domains)
% 26.2\% (180,468/689,492)
The second category is one of the most common tactics used by phishing attackers: exploiting squatted domains. These domains incorporate a modified version of a brand name in the domain or subdomain, closely mimicking legitimate brand domains to deceive users. For example, a phishing website targeting \cc{facebook}.\cc{com} might use a squatted domain such as \cc{faceb\{\textbf{o}\}ook}.\cc{com} to trick victims into believing they are accessing an authentic website.
\looseness=-1

To identify potential squatted domains, we employ the dnstwist tool~\cite{elceefdn84:online}, which generates domain name variations using various squatting techniques and widely used in previous works~\cite{CompareP38:online,sharma2024securing,gorboe2022detection,kaushik2021exploring}. 
We apply this tool to the top 200 most targeted brand names, which account for 90\% of the domains in our dataset. 
This process generates 765,444 possible squatted domains based on techniques such as adding extra characters to the domain name (\eg, \cc{facebook\textbf{0}}.\cc{com} from \cc{facebook}.\cc{com}), modifying a single bit in the domain name (\eg, \cc{fa\textbf{a}ebook}.\cc{com}), replacing characters with visually similar alternatives (\eg, \cc{faceb\textbf{0}ok}.\cc{com}, where \cc{o} is replaced with the number, \cc{0}), and adding hyphens or extra prefixes (\eg, \cc{face\textbf{-}book}.\cc{com} or \cc{\textbf{d}facebook}.\cc{com}).
Our analysis shows that 26.2\% of domains in our dataset are squatted domains.
%use squatted domains in their domain name or subdomain. 
% This finding highlights the significant role that squatting techniques play in phishing campaigns and emphasizes the need for robust detection strategies to mitigate this threat effectively.

% The second category is one of the most common tactic for phishing attackers is to exploit the squatted domains.
% The squatted domains contain a squatted brand name in a domain or subdomain. 
% For example, a phishing website targeting `facebook.com' can have a squatted domain of `faceb\textbf{o}ook.com'.

% To find a possible \texttt{squatted domain}, we utilize dnstwist~\cite{elceefdn84:online} tool to generate a \texttt{squatted domain} from a top 200 most used target brand names.
% A top 200 most used target brand names covers 90\% of domains from our dataset.
% From 200 domains, we generate 765,444 possible squatted domains.
% A \texttt{squatted domain} includes adding a character (\ie, facebook0.com from facebook.com), bitsquatting which modifies one bit in the domain name (\ie, faaebook.com from facebook.com), homoglyph which swaps character with similar looking digits (\ie, faccb0ok.com from facebook.com where o is replaced with 0), and adding hyphen or character (\ie, face-book.com, dfacebook.com).
% As a result, we find 26.2\% of domains from our dataset is using squatted domain in their domain name or subdomain.

% A maliciously registered domain has three key features within a lexical feature: contain benign brand name in a domain or subdomain (\ie, squatted domain), randomly generated domain, 

\PP{(3) Random-looking Algorithmic Domain}
The random appearance of algorithmically generated domains makes them hard to detect~\cite{UsingAno20:online,Automati98:online}. 
%Domains can also be generated algorithmically to appear randomized, making them harder to detect~\cite{UsingAno20:online,Automati98:online}. 
Attackers exploit this trend by capitalizing on users’ tendencies not to scrutinize domain names closely before clicking on links, even when the domain looks suspicious. 
% This strategy allows attackers to bypass both human vigilance and automated detection, highlighting the need for improved detection mechanisms that can effectively analyze algorithmically generated domains.
To find random-looking algorithmic domains, we follow the approach in~\cite{Automati98:online} by matching domains with English word lists.
%We utilize~\cite{SCOWLCus71:online} to find English words from the list of our domains, and then we eliminate domains that have English words.
We use ~\cite{SCOWLCus71:online}, a word list containing 108,687 words,  to identify domains that include any English words.
%This English word list~\cite{SCOWLCus71:online} contains 108,687 words.
We apply this process after removing the brand in the domain and squatted domains, leaving a total of  194,099 domains.%after removing domains that contain English words.
%, we are left with 194,099 domains.

As shown in~\autoref{tab:maliciously-registered-domain-categories}, a significant portion of domains (28.2\%) are random-looking algorithmic domains.
While such domains may appear suspicious to a human ~\cite{Automati98:online}, automated detection tools often struggle to classify them as malicious due to their lack of clear patterns or recognizable features.
% \DK{stat here}

\PP{(4) Bulk Registration of Domain}
%Other than brand names in domains, squatted domains, and randomly generated domains, we  suggest that bulk-registered domains can be maliciously registered.
%Within maliciously registered domains, attackers can register in bulk by registering multiple malicious domains simultaneously to maximize profit with minimal effort~\cite{hao2013understanding}. 
Attackers often  register many malicious domains simultaneously through bulk registration to maximize profits with minimal effort~\cite{hao2013understanding}. 
A phishing campaign can involve
%For example, an attacker can 
registering multiple domains at the same time and deploying multiple webpages with different domains.
%this way when 
Even if one domain is blocklisted, an attacker can rely on others to continue the attack.
%another source to execute an attack.
Our method to find bulk registered domains includes three conditions that must all be met: registered at the same time, registered through the same registrar, and domain names are similar (using Levenshtein distance~\cite{korczynski2018cybercrime}).
% \KL{Bulk registration category: registered in same time, same registrar, Levenshtein distance}

% Bulk registration is another form of attackers maximizing their results with minimal effort~\cite{hao2013understanding}.
% An attacker can register multiple domains at the same time and deploy multiple webpages with different domains, this way when one domain is blocklisted, an attacker has another source to execute an attack.
% To detect bulk registered domains, we set three conditions as introduced in prior work~\cite{maroofi2020comar}: registered at the same time, registered by the same registrar, and using Levenshtein distance shown in~\cite{korczynski2018cybercrime} to determine whether registered domains are similar.
% Interestingly, bulk-registered domains are often created simultaneously through the same registrar as shown in~\autoref{tab:registrar-in-bulk}, yet blocklists detect them at different times. 
% Even worse, Alibaba Cloud allows users to register domains in bulk~\cite{DomainNa2:online} and advertises with lower price~\cite{NewUserD82:online}.
% By registering multiple domains in bulk, attackers can sustain their activities, as some domains remain active even after others are detected. Some registrars allow bulk registrations, but they could proactively implement measures to prevent attackers from maliciously registering phishing domains in bulk.

\begin{table}[!t]
\caption{Top 10 Registrar in Bulk Registered Domains. \cc{Alibaba} stands out as the registrar associated with the highest number of bulk-registered domains.}
\label{tab:registrar-in-bulk}
\vspace{-10px}
\resizebox{0.9\linewidth}{!}{ 
\begin{NiceTabular}{r l r c}
\toprule
\multicolumn{1}{c}{\textbf{Rank}} & \multicolumn{1}{c}{\textbf{Registrar}} & \multicolumn{1}{c}{\textbf{\# of Domains}} & \multicolumn{1}{c}{\textbf{Country}}\\
\midrule
1 & ALIBABA SGP.${^\ast}$~\cite{AlibabaC38:online} & 4,180 (7.6\%)& CN\\
2 & Alibaba (Wanwang)${^\dagger}$~\cite{NewUserD82:online} & 2,599 (4.7\%)& CN \\
3 & SAV.COM~\cite{Sav54:online} & 2,093 (3.8\%) & US\\
4 & GoDaddy.com~\cite{GoDaddy24:online} & 1,845 (3.4\%)& US\\
5 & Gname.com Pte.~\cite{GNAMEBuy32:online} & 1,560 (2.8\%)& SGP\\
6 & Alibaba Cloud${^\ddagger}$~\cite{AlibabaC38:online} & 1,352 (2.5\%)& CN \\
7 & NameSilo~\cite{LowCostD67:online} & 1,285 (2.3\%)& US\\
8 & Network Solutions~\cite{DomainNa10:online} & 623 (1.1\%)& US\\
9 & Dynadot Inc~\cite{BuyaDoma33:online} & 618 (1.1\%)& US\\
10 & Aceville Pte.~\cite{DNSPod44:online} & 604 (1.1\%)& SGP\\
\midrule
Total &  & 54,787 (100\%)& - \\
\midrule
% \multicolumn{4}{l}{${^\ast}$: ALIBABA.COM SINGAPORE E-COMMERCE PRIVATE LIMITED} \\
\multicolumn{4}{l}{${^\ast}$: ALIBABA.COM SINGAPORE E-COMMERCE PRIVATE} \\
\multicolumn{4}{l}{${^\dagger}$: Alibaba Cloud Computing Co., Ltd. (Wanwang)} \\
\multicolumn{4}{l}{${^\ddagger}$: Alibaba Cloud Computing Ltd. d/b/a HiChina (www.net.cn)}
\end{NiceTabular}
}
\vspace{-15px}
\end{table}

% \begin{figure}[!t]
% \centering
%     \begin{subfigure}{0.4\textwidth}
%         \hspace{5px}\includegraphics[width=\linewidth]{fig/TLD_by_year_caption_v1.pdf}
%         \vspace{-12px}
%         % \caption{Top 10 TLD by Year All.}
%     \end{subfigure}
% \centering
%     \begin{subfigure}{0.16\textwidth}
%         \includegraphics[width=\linewidth,height=6.9em]{fig/TLD_by_year_all_v1.pdf}
%         \vspace{-10px}
%         % \caption{Top 10 TLD by Year All.}
%         \caption{Top 10 (All).}
%         \label{fig:TLD_by_year_all}
%     \end{subfigure}
%     \begin{subfigure}{0.15\textwidth}
%         \includegraphics[width=\linewidth]{fig/TLD_by_year_mal_v1.pdf}
%         \vspace{-10px}
%         % \caption{Top 10 TLD by Year (Malicious).}
%         \caption{Top 10 (Mal.).}
%         \label{fig:TLD_by_year_mal}
%     \end{subfigure}
%     \begin{subfigure}{0.15\textwidth}
%         \includegraphics[width=\linewidth]{fig/TLD_by_year_comp_v1.pdf}
%         \vspace{-10px}
%         % \caption{Top 10 TLD by Year (Comp.).}
%         \caption{Top 10 (Comp.).}
%         \label{fig:TLD_by_year_comp}
%     \end{subfigure}
%     \vspace{-10px}
%     \caption{Top 10 TLD by Year. While \cc{.com} is the most used, \cc{.shop}, \cc{.cn} increase over the years.\KL{make lines thicker, try different colors}}
%     \label{fig:top10_TLD_by_year}
%     % \vspace{-10px}
% \end{figure}

\begin{table*}[t]
\caption{Top 10 Targeted Brands. 
Popular brands (\eg, USPS, OZON, Instagram) predominantly utilize \textcolor{magenta}{\textbf{.top}}, \textcolor{magenta}{\textbf{.tk}}, \textcolor{magenta}{\textbf{.ml}} than the origin of its brand (\eg, \textcolor{dkgreen}{\textbf{.com}}, \textcolor{dkgreen}{\textbf{.ru}}). 
}\vspace{-10px}
\label{tab:impersonated_brand}
\resizebox{0.98\linewidth}{!}{ 
% \begin{NiceTabular}{l c r r r l r r l r r}
% \begin{NiceTabular}{l c r r r l r r l r r}
\begin{tabular}{l c r r r l r r l r r}
\toprule
\multirow{2}{*}{\parbox{1cm}{\centering\textbf{Brand}}} & \multirow{2}{*}{\centering \textbf{Country}} & \multirow{2}{*}{\centering \textbf{Total}} & 
% \multicolumn{1}{c}{\multirow{2}{*}{\textbf{Malicious\\ domains}}} 
% \multicolumn{1}{c}{\multirow{2}{*}{\makecell{{\textbf{Malicious}}\\{\textbf{Domains}}}}}
\makecell{\multirow{2}{*}{\textbf{Malicious}} \\ \multirow{2}{*}{\textbf{Domains}}}
& 
% \multicolumn{1}{c}{\multirow{2}{*}{\textbf{Compromised}}}
\multicolumn{1}{c}{\multirow{2}{*}{\makecell{{\textbf{Compromised}}\\{\textbf{Domains}}}}}
& \multicolumn{3}{c}{\textbf{\makecell{Malicious Domains}}} & \multicolumn{3}{c}{\textbf{Compromised Domains}}\\
% \cmidrule(l){6-11}
\cmidrule(lr){6-8}\cmidrule(lr){9-11}
&
& 
& 
&
& \multicolumn{1}{c}{\textbf{TLD${^\ast}$}} & \multicolumn{1}{c}{\textbf{TLD Count}} & \multicolumn{1}{c}{\textbf{Unique${^\S}$}} & \multicolumn{1}{c}{\textbf{TLD${^\ast}$}} & \multicolumn{1}{c}{\textbf{TLD Count}} & \multicolumn{1}{c}{\textbf{Unique${^\S}$}}\\
\midrule
Facebook & US & 66,700 & 38,817 (58.2\%) & 27,227 (40.8\%) & .com & 11,485 (29.6\%) & 439 & .com & 10,764 (39.5\%) & 324 \\
USPS & US & 41,691 & 37,533 (90.0\%) & 4,109 \xspace\xspace(9.9\%) & \cellcolor{bubblegum}\textbf{.top} & 15,489 (41.3\%) & 259 & \cellcolor{bubblegum}\textbf{.top} & 1,835 (44.7\%) & 153 \\
Microsoft & US & 26,717 & 13,681 (51.2\%) & 12,438 (46.6\%) & .com & 5,759 (42.1\%) & 449 & .com & 6,358 (51.1\%) & 371 \\
DHL & GER${^\dagger}$ & 23,539 & 15,784 (67.1\%) & 7,277 (30.9\%) & .com & 5,741 (36.4\%) & 451 & .com & 3,612 (49.6\%) & 322 \\
OZON & RUS${^\dagger}$ & 18,513 & 10,248 (55.4\%) & 8,465 (45.7\%) & \cellcolor{bubblegum}\textbf{.tk} & 4,549 (44.4\%) & 34 & \cellcolor{bubblegum}\textbf{.tk} & 4,600 (54.3\%) & 17 \\
WhatsApp & US & 11,521 & 8,264 (71.7\%) & 3,163 (27.5\%) & .com & 2,363 (28.6\%) & 162 & .com & 1,198 (37.9\%) & 104 \\
Apple & US & 11,253 & 8,942 (79.5\%) & 2,056 (18.3\%) & .com & 3,385 (37.9\%) & 234 & .com & 918 (44.6\%) & 138 \\
Instagram & US & 11,181 & 7,337 (65.6\%) & 3,681 (32.9\%) & \cellcolor{bubblegum}\textbf{.ml} & 1,482 (20.2\%) & 212 & .com & 982 (26.7\%) & 165 \\
Naver & KOR${^\dagger}$ & 11,030 & 7,207 (65.3\%) & 3,725 (33.8\%) & .com & 2,506 (34.8\%) & 269 & .com & 1,549 (41.6\%) & 195 \\
Amazon & US & 9,473 & 7,390 (78.0\%) & 2,000 (21.1\%) & .com & 2,086 (28.2\%) & 192 & .com & 996 (49.8\%) & 110 \\
\midrule
\multicolumn{11}{l}{${^\ast}$: Most common TLD in brands. ${^\S}$: \# of unique TLD in brand. ${^\dagger}$: GER: Germany, RUS: Russia, KOR: Korea.}
% \multicolumn{11}{l}{${^\ddagger}$: Using other TLD (\eg, \textcolor{magenta}{\textbf{.top}}, \textcolor{magenta}{\textbf{.tk}}, \textcolor{magenta}{\textbf{.ml}}) than the origin of its brand (\eg, \textcolor{dkgreen}{\textbf{.com}}, \textcolor{dkgreen}{\textbf{.ru}})} 
\end{tabular}
}
\end{table*}

\begin{figure*}[!t]
\vspace{-5px}
\centering
        \begin{subfigure}{0.9\textwidth}
            % \hspace{5px}
            \includegraphics[width=\linewidth]{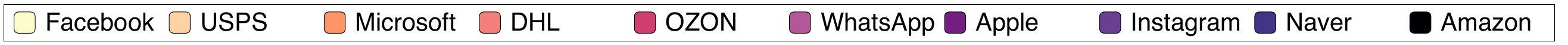}
            % \vspace{-12px}
            % \caption{Top 10 TLD by Year All.}
        \end{subfigure}
\centering
    \begin{subfigure}{0.32\textwidth}
        \includegraphics[width=\linewidth,height=9em]{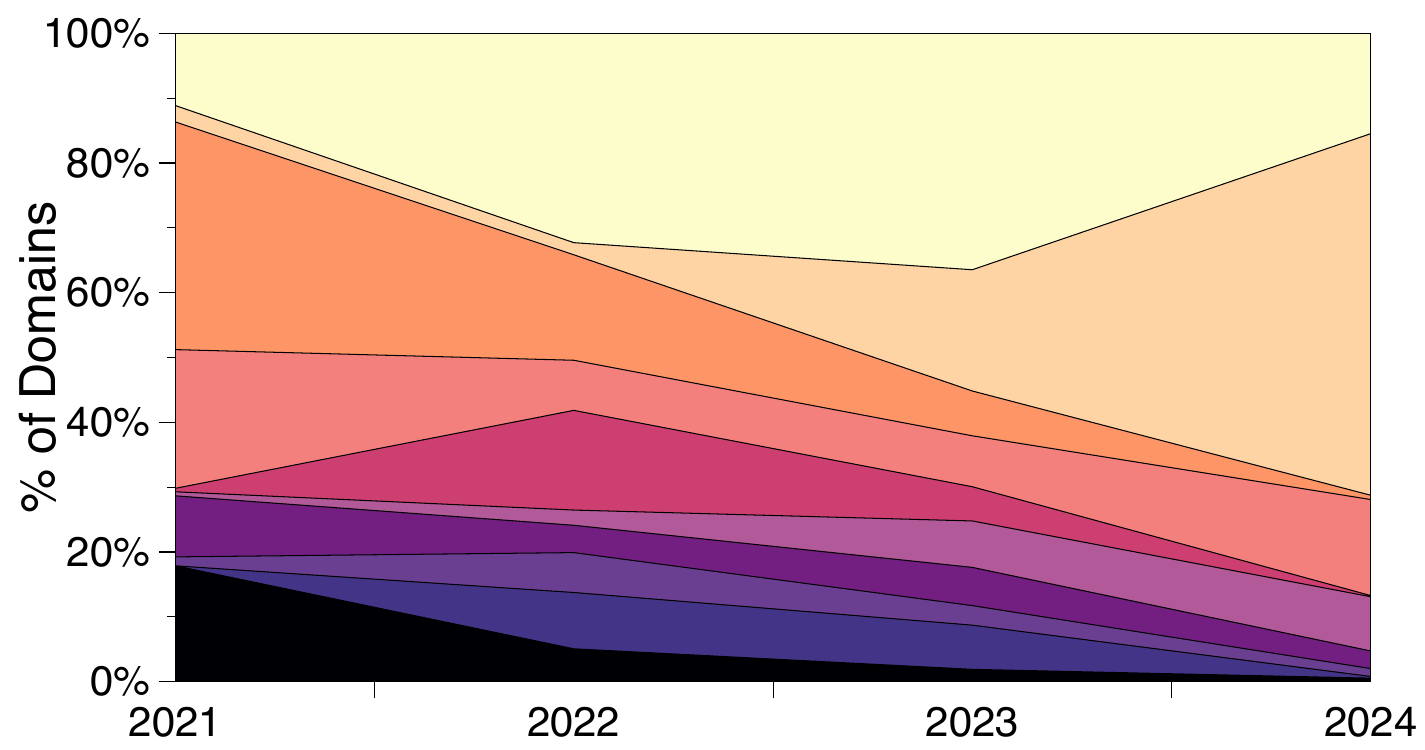}
        \vspace{-10px}
        \caption{Top 10 Brand by Year All.}
        \label{fig:Top10_brand_by_year_all}
    \end{subfigure}
    \begin{subfigure}{0.32\textwidth}
        \includegraphics[width=\linewidth]{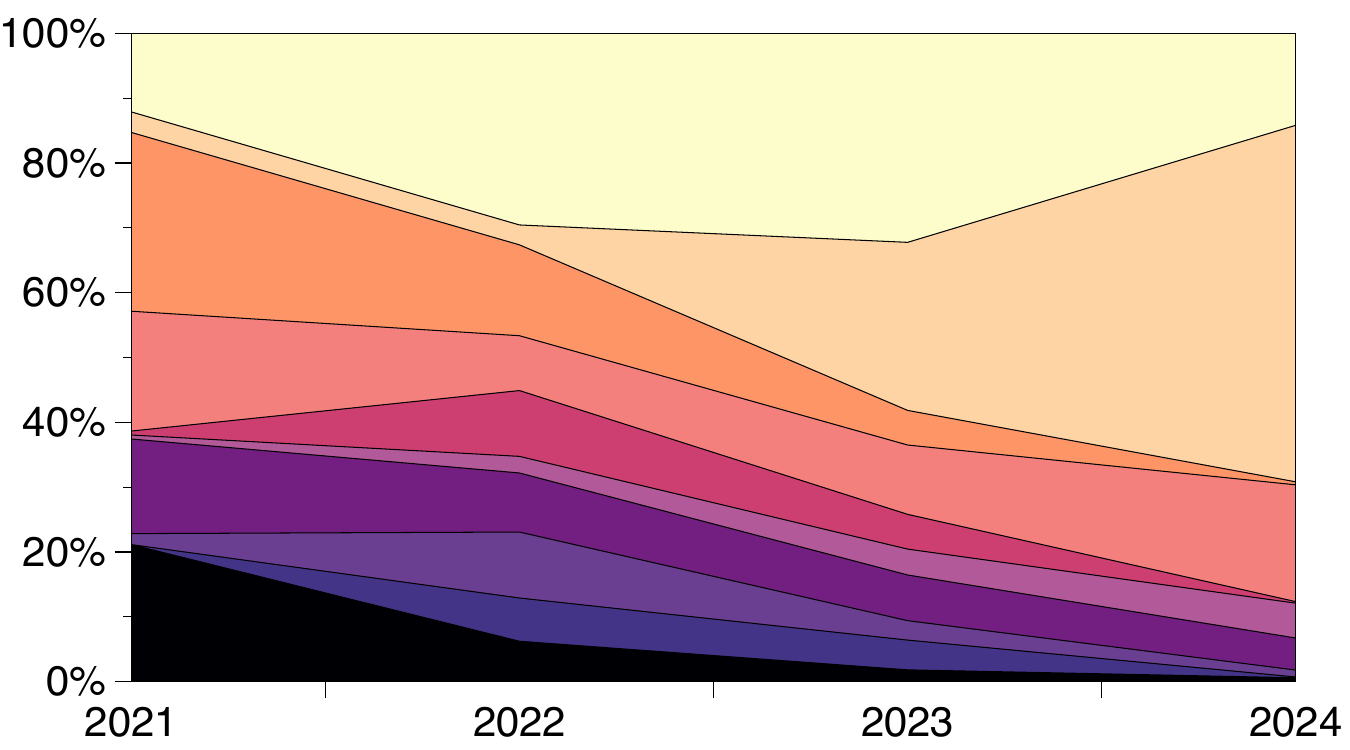}
        \vspace{-10px}
        \caption{Top 10 Brand by Year (Malicious).}
        \label{fig:Top10_brand_by_year_mal}
    \end{subfigure}
    \begin{subfigure}{0.32\textwidth}
        \includegraphics[width=\linewidth]{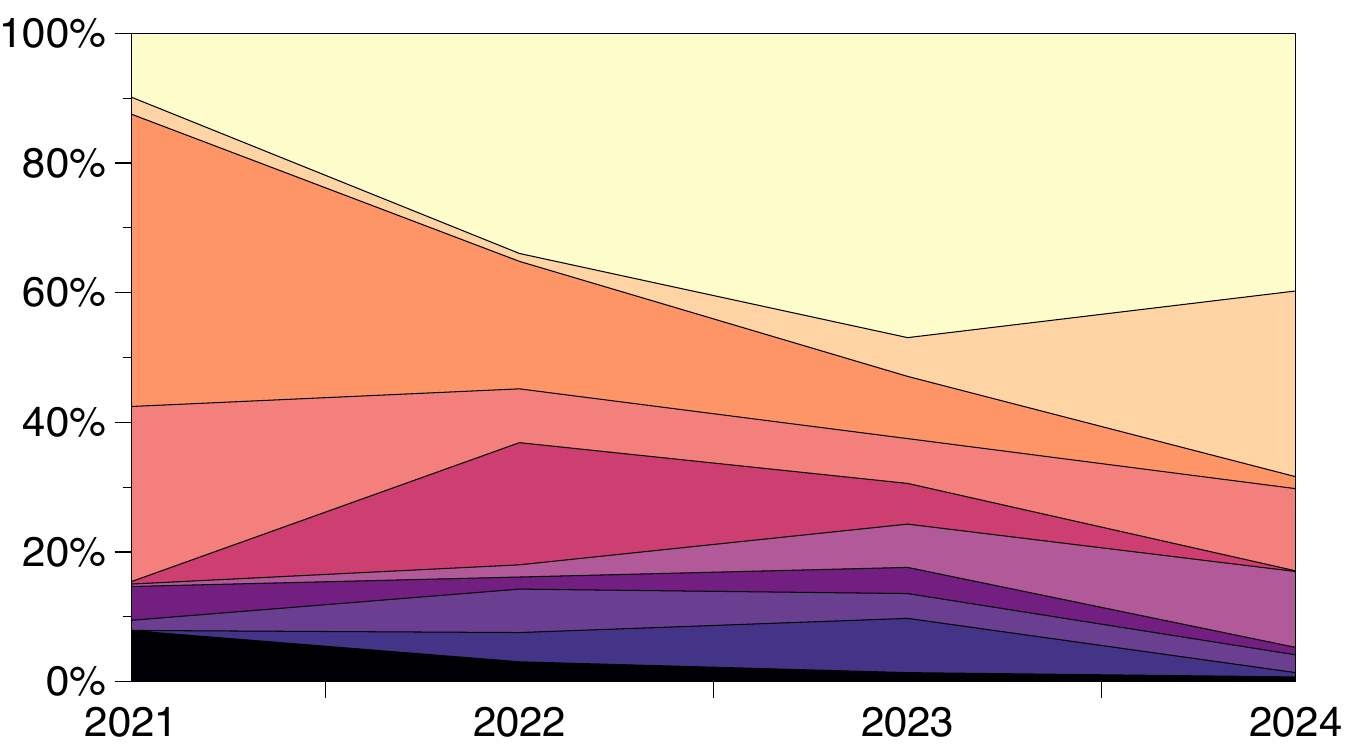}
        \vspace{-10px}
        \caption{Top 10 Brand by Year (Comp.).}
        \label{fig:Top10_brand_by_year_comp}
    \end{subfigure}
    \vspace{-10px}
    \caption{Top 10 Brand by Year. USPS increases dramatically from 2022 to 2024, specifically in maliciously registered domains. On the other hand, Microsoft decreases in all domains, DHL increases in maliciously registered domains but decreases in the compromised domains.}
    \label{fig:top10_brand_by_year}
    \vspace{-10px}
\end{figure*}

% \DK{stat here: coverage}
Bulk-registered domains, often created simultaneously through the same registrar, account for 7.9\% of the domains in our dataset, as shown in~\autoref{tab:registrar-in-bulk}. While this percentage represents a smaller subset of the dataset, it carries significant implications.
\looseness=-1
% Identifying bulk-registered domains provides an opportunity to mitigate phishing attacks by restricting or disallowing bulk registrations, thereby disrupting attackers' ability to profit from this tactic.

% \DK{screenshot reference is missing.}
% To better understand this phenomenon, we examine which registrars facilitate bulk registration and identify those responsible for the highest number of bulk-registered domains, as shown in~\autoref{tab:registrar-in-bulk}. 
% Notably, \cc{Alibaba} \cc{Cloud}~\cite{AlibabaC38:online} frequently appears as a registrar for such domains, offering bulk registration services~\cite{DomainNa2:online} and even further, promoting them with discounted pricing~\cite{NewUserD82:online} as illustrated in \RM{??}~\autoref{fig:alibaba_bulk_registration}. 

Notably, \cc{Alibaba} \cc{Cloud}\cite{AlibabaC38:online} frequently serves as a registrar for these domains, offering bulk registration services\cite{DomainNa2:online}. 
Furthermore, it actively promotes bulk registrations through discounted pricing~\cite{NewUserD82:online}, as illustrated in~\autoref{apx:bulk_registration}.% ~\autoref{fig:alibaba_bulk_registration}. 
This combination of bulk registration functionality and discounted pricing likely lowers the barrier for registering multiple domains, making it an attractive option for attackers.
This practice enables attackers to sustain their operations by registering multiple domains in bulk, ensuring that some remain active even after others are detected. 
%While some registrars allow bulk registrations, adopting proactive measures, such as stricter verification or limits on bulk purchases, could significantly reduce the malicious use of bulk-registered domains.
Some registrars adopt proactive measures, such as stricter verification or limits on bulk purchases, which significantly reduce the malicious use of bulk-registered domains.

\looseness=-1

\PP{Manual Validation}
We randomly select 1,000 domains from our identified maliciously registered domains.
Then, we manually validate our method of identifying maliciously registered domains by examining the contents of the phishing domains.
% and DNS records.
Specifically, we utilize historical data from the Wayback Machine~\cite{WaybackM89:online} to identify domains that either lack historical snapshots or display content designed to mimic legitimate webpages.

Our analysis reveals that 72.3\% of the examined domains do not have any historical data in the Wayback Machine. Among the remaining 27.7\%, 14.8\% domains redirect to error pages, while the remaining 12.9\% of domains host malicious content pages. 
% These findings provide additional validation of our approach and highlight the characteristics of maliciously registered phishing domains.

% \DK{fill this..}

% This is based on our assumption that maliciously registered domains either 
% lack historical snapshots or exhibit content designed to mimic legitimate webpages. 

\rtbox{
\textbf{Takeaway:} 
We combined the existing method with our new method of identifying maliciously registered domains.
Maliciously registered domains are over half of phishing domains (66.1\%). 
% \RM{I dont understand the previous sentence. Do you mean Over half of the phishing domains are registered by attackers?}\KL{yes}
Phishing attackers often exploit bulk registration services, such as those offered by \cc{Alibaba} \cc{Cloud}. Notably, among registrars that provide bulk registration, \cc{Alibaba} emerges as the most frequently abused platform for registering domains in bulk.
}

\section{Characteristics of Maliciously registered Domains}
\label{sec:mal-reg-char}

% \begin{figure*}[h]
% \centering
%     \begin{subfigure}{0.32\textwidth}
%         \includegraphics[width=\linewidth]{fig/TLD_by_year_all_02.pdf}
%         \vspace{-20px}
%         \caption{Top 10 TLD by Year All.}
%         \label{fig:TLD_by_year_all}
%     \end{subfigure}
%     \begin{subfigure}{0.32\textwidth}
%         \includegraphics[width=\linewidth]{fig/TLD_by_year_mal_02.pdf}
%         \vspace{-20px}
%         \caption{Top 10 TLD by Year (Malicious).}
%         \label{fig:TLD_by_year_mal}
%     \end{subfigure}
%     \begin{subfigure}{0.32\textwidth}
%         \includegraphics[width=\linewidth]{fig/TLD_by_year_comp_02.pdf}
%         \vspace{-20px}
%         \caption{Top 10 TLD by Year (Comp.).}
%         \label{fig:TLD_by_year_comp}
%     \end{subfigure}
%     \vspace{-10px}
%     \caption{Top 10 TLD by Year. While \cc{.com} is the most used, \cc{.shop}, \cc{.cn} increase over the years.}
%     \label{fig:top10_TLD_by_year}
%     \vspace{-10px}
% \end{figure*}

% \input{tables/top10_tld_year}

%maliciously registered domains,
We analyze DNS components of maliciously registered domains, including their targeted brands, TLDs, and DNS records, to gain insights into their characteristics.

\subsection{Targeted Brand}
We utilize target brand information from the APWG dataset.
In our analysis, we identify a diverse range of targeted brands, with \cc{Facebook} standing out as the most targeted brand, followed by \cc{USPS} as shown in~\autoref{tab:impersonated_brand}. 
These two brands alone account for a significant portion with 15.5\% (108,391 out of 697,237) of phishing domains, reflecting their widespread recognition and trust among users. 

\begin{table*}[!t]
\caption{Top 10 TLDs by Year. New-gTLD with lower registration costs is widely used in maliciously registered (\textcolor{magenta}{\textbf{.top}}, \textcolor{magenta}{\textbf{.xyz}}, \textcolor{magenta}{\textbf{.shop}}, \textcolor{magenta}{\textbf{.online}}) than compromised domains (\textcolor{dkgreen}{\textbf{.net}}, \textcolor{dkgreen}{\textbf{.info}}). Freenom usage decreases and (\eg, \textcolor{magenta}{\textbf{.cn}} increased in 2024 where \textcolor{dkgreen}{\textbf{.tk}} and \textcolor{dkgreen}{\textbf{.ml}} decreased)}
\label{tab:tld_by_year}
\vspace{-10px}
\resizebox{1\linewidth}{!}{ 
% \begin{NiceTabular}{lrrrrrrrrrrrrcc}
\begin{tabular}{lrrrrrrrrrrrrcc}
\toprule
% \multicolumn{1}{c}{\multirow{2}{*}{\textbf{Total}}}
\multicolumn{1}{c}{\multirow{2}{*}{\textbf{TLD}}} & \multicolumn{1}{c}{\multirow{2}{*}{\textbf{Total}}}
& \multicolumn{5}{c}{\textbf{\makecell[l]{Maliciously Registered Domains}}} & \multicolumn{5}{c}{\textbf{Compromised Domains}} & \multicolumn{1}{c}{\multirow{2}{20px}{\textbf{Price\\(USD)${^\ast}$}}} & \multicolumn{1}{c}{\multirow{2}{*}{\textbf{Types${^\dagger}$}}} & \multicolumn{1}{c}{\multirow{2}{*}{\textbf{Freenom}}}\\
% \cmidrule(l){3-12}
\cmidrule(lr){3-7}\cmidrule(lr){8-12}
% \multicolumn{1}{c}{\textbf{TLD}} & \multicolumn{1}{c}{\textbf{Total}}  
&
& \multicolumn{1}{c}{\textbf{2021}} & \multicolumn{1}{c}{\textbf{2022}} & \multicolumn{1}{c}{\textbf{2023}} & \multicolumn{1}{c}{\textbf{2024}} & \multicolumn{1}{c}{\textbf{Total}} & \multicolumn{1}{c}{\textbf{2021}} & \multicolumn{1}{c}{\textbf{2022}} & \multicolumn{1}{c}{\textbf{2023}} & \multicolumn{1}{c}{\textbf{2024}} & \multicolumn{1}{c}{\textbf{Total}} & & & \\
% \multicolumn{1}{c}{\textbf{Price}} & \multicolumn{1}{c}{\textbf{Types}} & \multicolumn{1}{c}{\textbf{Freenom}}\\
\midrule
.com & 218,267 & 19,359 & 53,795 & 42,376 & 26,949 & 142,479 & 10,837 & 30,051 & 20,813 & 14,087 & 75,788 & \$6 & gTLD & No \\
.top & 84,686 & 2,735 & 12,792 & 19,519 & 18,968 & 54,014 & 3,005 & 10,612 & 13,721 & 3,334 & 30,672 & \textbf{\$1} & new gTLD & No \\
.xyz & 37,698 & 3,545 & 10,520 & 5,396 & 3,624 & 23,085 & 2,985 & 8,310 & 2,732 & 586 & 14,613 & \textbf{\$1} & new gTLD & No \\
.shop & 30,065 & 764 & 2,790 & 5,905 & 11,081 & 20,540 & 628 & 1,532 & 2,550 & 4,815 & 9,525 & \textbf{\$1} & new gTLD & No \\
.cn & 24,708 & 1,639 & 9,234 & 2,045 & 7,060 & \cellcolor{bubblegum}\textbf{19,978} & 471 & 798 & 1,582 & 1,879 & \cellcolor{lightgreen}\textbf{4,730} & \$5 & ccTLD & No \\
.tk & 22,453 & 779 & 10,931 & 1,802 & \cellcolor{lightgreen}\textbf{52} & 13,564 & 213 & 7,001 & 1,623 & \cellcolor{lightgreen}\textbf{52} & 8,889 & \$7 & ccTLD & \cellcolor{light-gray}\textbf{Yes} \\
.online & 14,409 &679 & 3,609 & 3,084 & 1,271 & 8,643 & 254 & 3,123 & 1,741 & 648 & 5,766 & \textbf{\$1} & new gTLD & No \\
.ml & 14,154 &726 & 8,868 & 526 & \cellcolor{lightgreen}\textbf{5} & 10,125 & 251 & 3,388 & 390 & \cellcolor{lightgreen}\textbf{0} & 4,029 &\$12 & ccTLD & \cellcolor{light-gray}\textbf{Yes} \\
.net & 12,672 &1,302 & 3,217 & 2,309 & 1,260 & 8,088 & 691 & 1,987 & 1,233 & 673 & 4,584 &  \$10 & gTLD & No \\
.info & 10,619 &974 & 2,891 & 2,182 & 2,076 & 8,123 &  266 & 934 & 843 & 453 & 2,496 & \$2 & gTLD & No \\
\midrule
\multicolumn{15}{l}{${^\ast}$: Cost to register a domain in each TLD~\cite{CompareP38:online}. ${^\dagger}$: gTLD vs. ccTLD. Note that years 2021 and 2024 are not 12 months.}
\end{tabular}
}
\end{table*}

\begin{figure*}[!t]
    \vspace{-5px}
    \centering
        \begin{subfigure}{0.8\textwidth}
            % \hspace{5px}
            \includegraphics[width=\linewidth]{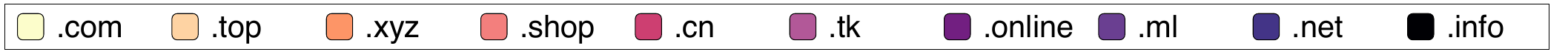}
            % \vspace{-12px}
            % \caption{Top 10 TLD by Year All.}
        \end{subfigure}
    \centering
    \begin{subfigure}{0.33\textwidth}
        \includegraphics[width=\linewidth,height=9em]{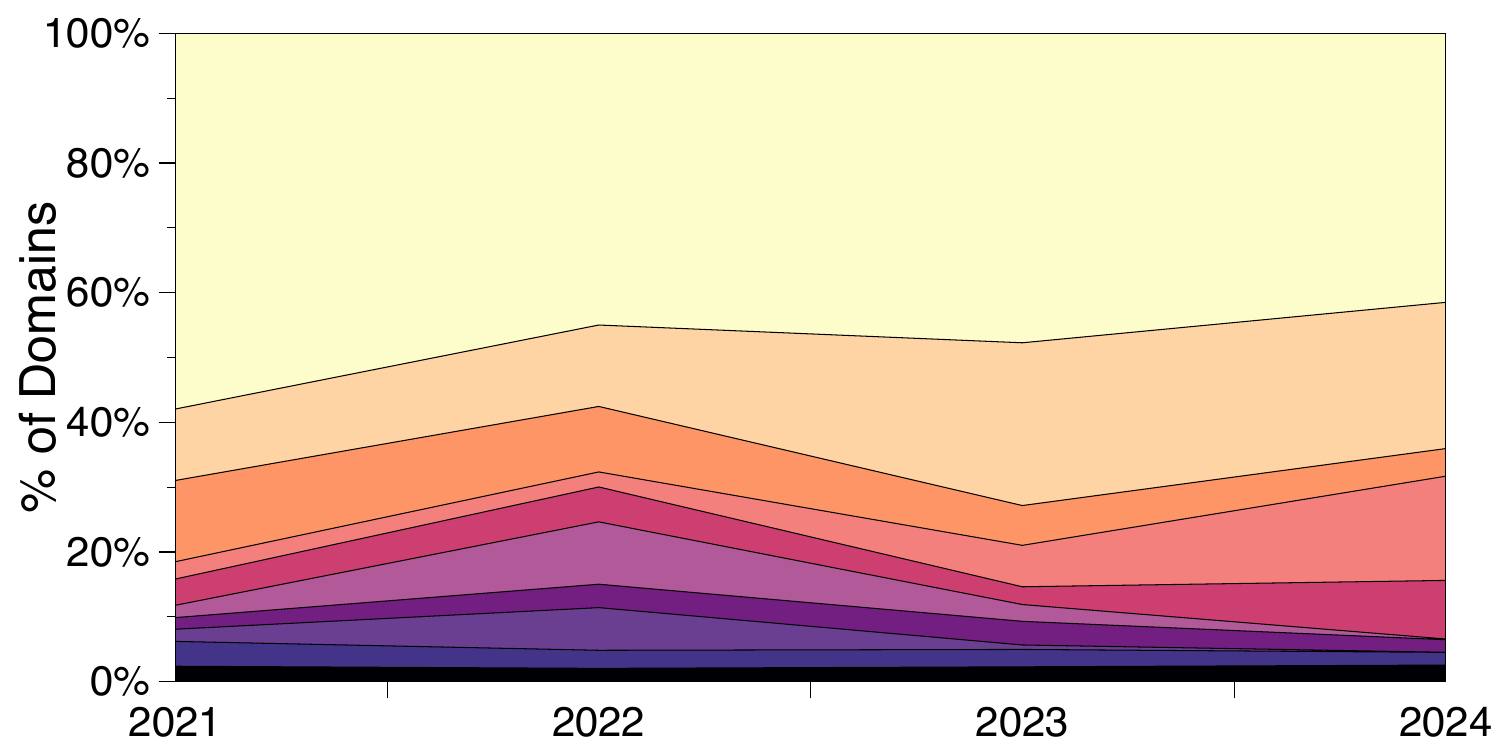}
        \vspace{-10px}
        \caption{Top 10 TLD by Year All.}
        \label{fig:TLD_by_year_all_stacked}
    \end{subfigure}
    \begin{subfigure}{0.32\textwidth}
        \includegraphics[width=\linewidth]{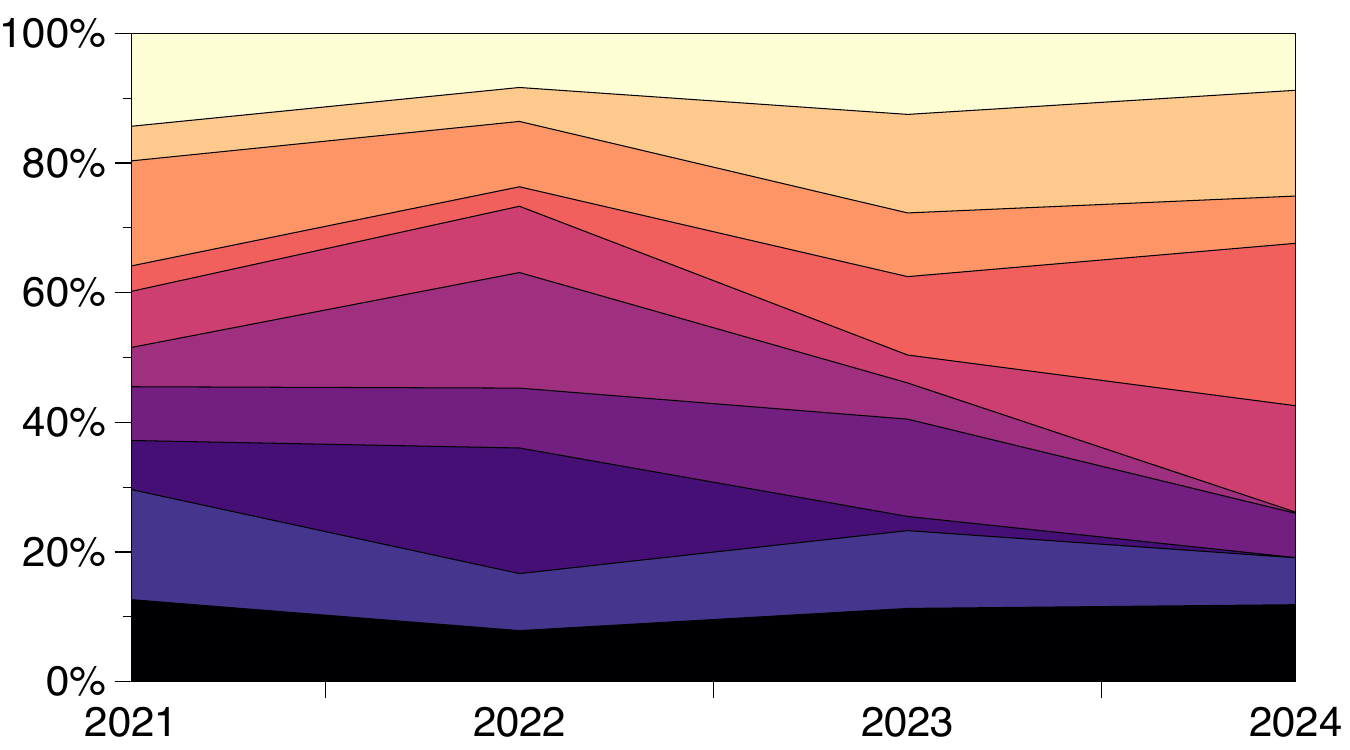}
        \vspace{-10px}
        \caption{Top 10 TLD by Year (Malicious).}
        \label{fig:TLD_by_year_mal_stacked}
    \end{subfigure}
     \begin{subfigure}{0.32\textwidth}
        \includegraphics[width=\linewidth]{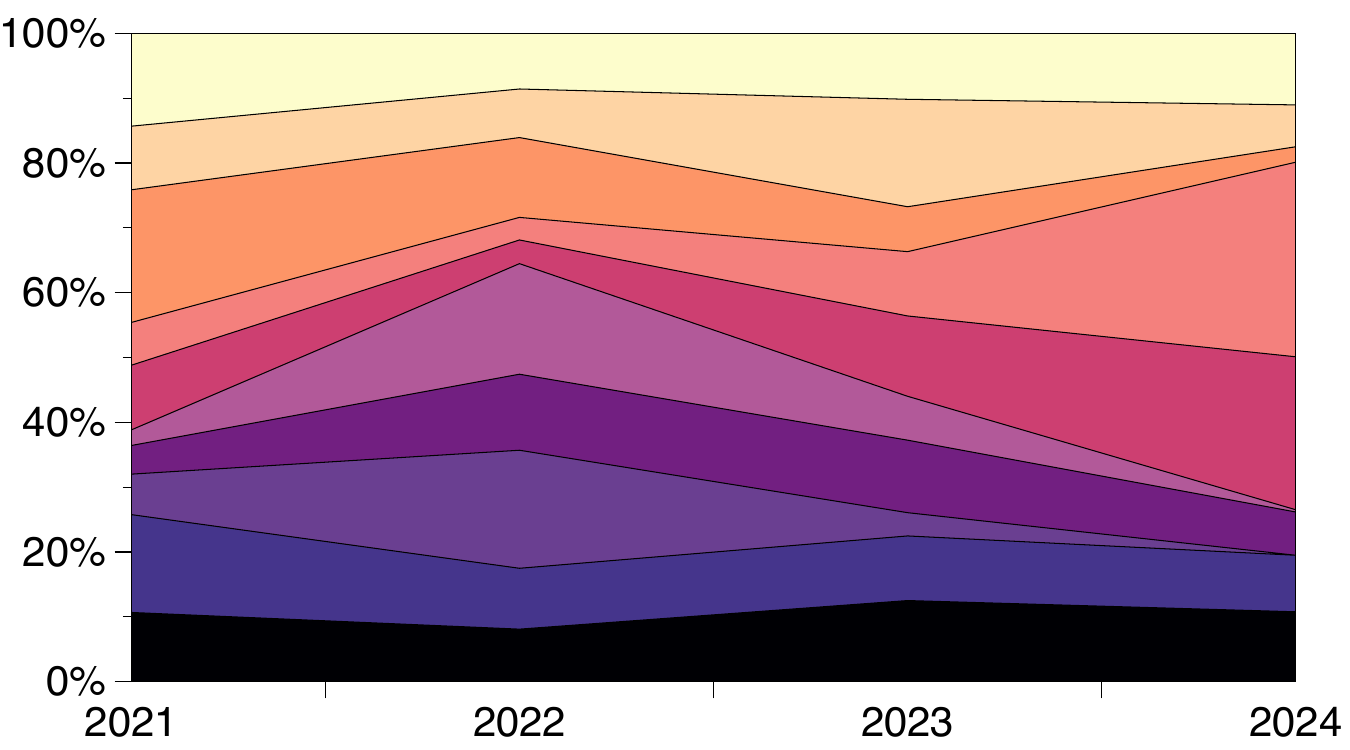}
        \vspace{-10px}
        \caption{Top 10 TLD by Year (Comp.).}
        \label{fig:TLD_by_year_comp_stacked}
    \end{subfigure}
    \vspace{-10px}
    \caption{Top 10 TLD by Year. While \cc{.com} is the most used, \cc{.shop}, \cc{.cn} increase over the years.}
    \label{fig:top10_TLD_by_year}
    \vspace{-10px}
\end{figure*}

% \PP{rend of Targeted Brand}
% \KL{Need to check. general trend here.}
As shown in ~\autoref{fig:top10_brand_by_year}, a clear trend emerges among popular targeted brands. 
Notably, \cc{USPS} is the second most targeted brand, accounting for 6.0\% of phishing domains. 
Interestingly, while \cc{USPS}-targeted domains were minimal in 2021 and 2022, there has been a dramatic increase since 2023. 
This finding aligns with previous reports on phishing domain trends~\cite{Phishing18:online}. 
Conversely, \cc{Microsoft} shows an overall decline in targeting, with a more pronounced decrease observed in compromised domains, as illustrated in ~\autoref{fig:Top10_brand_by_year_comp}. 
Additionally, \cc{DHL}-targeted domains demonstrate an increasing trend in maliciously registered domains over the years, while showing a decline in compromised domains. 

\begin{figure*}[!t]
    \centering
    \includegraphics[width=.98\linewidth]{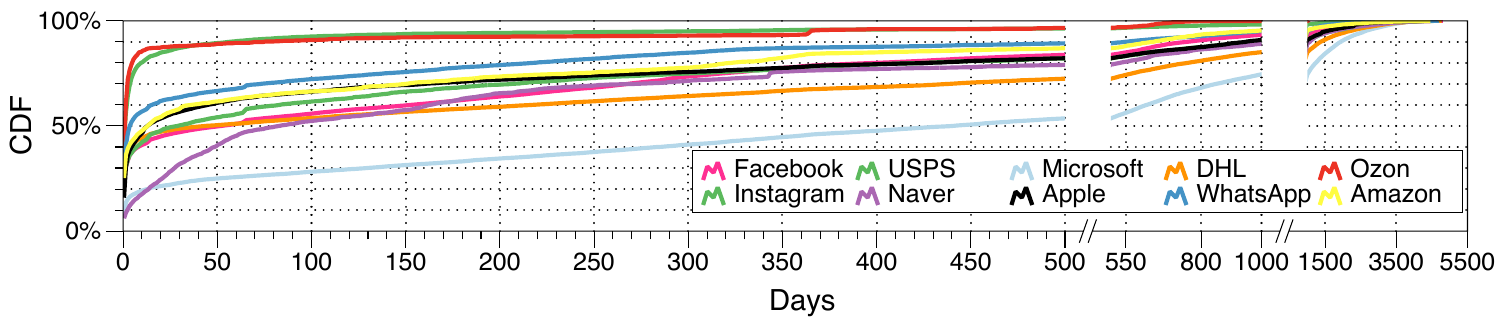}
    \vspace{-15px}
    \caption{Days Between Registration and Detection by Top 10 Brand.
    }
    \label{fig:registration_by_brand}
    \vspace{-10px}
\end{figure*}

\subsection{Top-level domain (TLD)}
\label{sec:tld_characteristics}
% \KL{add comparison between all TLD vs maliciously-registered domains' TLD}
% TLD can play a role when registering a domain.
% Whether attacker aims to register through cheaper TLD, or maximize impersonation with brand (\ie, using same TLD as benign TLD is using).
% In one of the phishing report~\cite{Phishing18:online}, Freenom TLD was one of the largest player for phishing attackers to register malicious domains since they provide registration for free.
% However since a reports~\cite{} discovered that phishing attackers abuse this functionality to register malicious domains, Freenom TLD do not provide registration for free from early 2023.
% A report from Interisle~\cite{Phishing18:online} show that after Freenom stop providing registration for free, phishing domains in ccTLD has increased.
% We also want to compare how our analysis hold up with this result by analyzing TLD usage in phishing domains specific focus on maliciously registered domains.

We investigate the use of TLDs in phishing domains %utilize various TLDs 
and assess whether certain TLDs are disproportionately abused. 
Our analysis considers the varying registration costs across TLDs, which may influence attackers' choices and strategies.
% \DK{Analysis Plan} \KL{What we are analyzing in this section}

\PP{Motivation}
TLD choice plays a significant role in domain registration for phishing attacks. Attackers may opt for cheaper TLDs to minimize costs or strategically use the same TLD as the targeted brand to enhance impersonation (\eg, using \cc{.com} for brands that also use \cc{.com}). 
According to a phishing report~\cite{Phishing18:online}, Freenom TLDs were among the most commonly exploited by phishing attackers, as they offered free registrations. 
However, after reports revealed widespread abuse of this functionality for malicious domain registration, Freenom ceased offering free registrations in early 2023.
% \KH{Same policy change described multiple times with similar context \color{red} (B)}
Moreover, a subsequent report by Interisle~\cite{Phishing18:online} noted a shift, with phishing domains in ccTLDs increasing after Freenom's policy change. To examine whether our findings align with this trend, we analyze TLD usage in phishing domains, focusing specifically on maliciously registered domains to uncover patterns and their implications.

\PP{Result: Trend of TLD}
~\autoref{tab:tld_by_year} highlights significant trends in TLD usage across phishing domains, illustrating attackers’ preferences and the influence of policy changes. 
The \cc{.com} TLD dominates the landscape with 218,267 (31.3\% out of 697,237) total phishing domains, likely due to its credibility and widespread familiarity, which enhance its effectiveness for deception. 
Low-cost new gTLDs, such as \cc{.top} and \cc{.shop}, become prominent in our result, with 84,686 (12.1\%) and 37,698 (5.4\%) domains, respectively, reflecting attackers’ preference for inexpensive and lenient TLDs. 
As shown in~\autoref{tab:tld_by_year}, the lower registration costs of new gTLDs (with prices as low as \$1 in our dataset) may contribute to their increased exploitation by phishing domains.
% \RM{USD?}\KL{yes} 
% \DK{add domain specific price}

Freenom TLDs (\eg, \cc{.tk}) are heavily exploited in earlier years, but seen a dramatic decline, dropping from 10,931 domains in 2022 to merely 52 domains in 2024, after Freenom discontinued free registrations in 2023.
% \KH{Same policy change described multiple times with similar context \color{red} (B)}
This finding aligns with a previous report~\cite{Phishing18:online}.
% This shift underscores how stricter policies can significantly reduce abuse. 
Additionally, as shown in~\autoref{fig:top10_TLD_by_year}, the growing presence of \cc{.cn}, from 764 in 2021 to 7,060 in 2024 domains, signals a strategic adaptation by attackers to target TLDs with potentially weaker enforcement mechanisms~\cite{Phishing18:online}.
\looseness=-1

As illustrated in ~\autoref{fig:top10_TLD_by_year}, there is a notable increasing trend in the use of new gTLDs, particularly \cc{.top} and \cc{.shop}. 
Interestingly, the use of \cc{.top} in maliciously registered domains has steadily increased over the years, while its usage in compromised domains shows a decline in 2024. 
In contrast, \cc{.shop} demonstrates a consistent increase in usage across both maliciously registered and compromised domains.
% As illustrated in~\autoref{fig:top10_TLD_by_year}, we can also see the increasing trend of new gTLDs specifically \cc{.top}, \cc{.shop}.
% Interestingly, the usage of \cc{.top} in maliciously registered domains increase over years however in compromised domains, it decreases in 2024.
% On the other hand, \cc{.shop} shows constant increase in between maliciously registered and compromised domains.
% \DK{what ground?}
% \DK{may need a line graph over time.}
% \DK{deeper analysis required}\KL{added trend result}

\rtbox{
\textbf{Takeaway:} 
%Our TLD analysis reveals that 
Phishing domains often exploit new gTLDs due to their lower registration costs. Notably, when \cc{Freenom} discontinued offering free domain registrations, the usage of \cc{.cn} domains increased concurrently. 
%Additionally, 
Certain new gTLDs, such as \cc{.shop}, exhibit distinct trends between maliciously registered domains and compromised domains, highlighting different attack strategies.
%employed by attackers.
% \DK{takeaway box (.cn becomes popular and freenom TLD is not used)}
}

\PP{Using Different TLD than Original Brand Domain}
\label{sec:using_diff_tld}
Phishing domains do not always use the same TLD as their original domains. 
For instance, 
%while \cc{Facebook}’s legitimate domain uses \cc{.com}, 
phishing attackers often register \cc{Facebook}-targeted domains using alternative TLDs such as \cc{.top}, rather than \cc{.com} that used by \cc{Facebook}. 
Similarly, \cc{USPS}, the second popular targeted brand in our analysis, is frequently targeted using \cc{.top} domains instead of the brand’s original \cc{.com}. 
Another example is \cc{OZON}, ranked as the 5th most targeted brand, with 49.42\% of its phishing domains registered under the \cc{.tk} instead of its original \cc{.ru}. 
Both \cc{.top} and \cc{.tk} are significantly cheaper than \cc{.com} for registration, with \cc{.tk} previously offered for free by Freenom until January 2023. 
Interestingly, both targeted brands \cc{USPS} and \cc{OZON} have the quickest detected time by blocklists as shown in~\autoref{fig:registration_by_brand}.
We will discuss detection time across different brands in~\autoref{sec:registration_detection}.

Another noteworthy observation from ~\autoref{tab:impersonated_brand} is that 44.39\% of \cc{OZON}-targeted domains are registered under \cc{.tk}, which is significantly more popular than any other brand. Additionally, \cc{OZON}-targeted domains exhibit the smallest number of unique TLDs (34) among the top 10 brands.
%, with only 34 unique TLDs. 
Furthermore, as shown in ~\autoref{fig:top10_brand_by_year}, \cc{OZON} demonstrates a decline in phishing activity over time. 
These findings suggest that attackers targeting \cc{OZON} often prefer low-cost TLDs, such as those offered by Freenom, to minimize costs and maximize the scalability of their phishing campaigns.

\rtbox{
\textbf{Takeaway:} 
Phishing attackers prefer low-cost TLDs like \cc{.top} and \cc{.tk} to target brands such as \cc{USPS} and \cc{OZON}, with \cc{OZON} relying on \cc{.tk} for 44.39\% of its phishing domains. These brands also show the fastest detection times by blocklists, highlighting the importance of monitoring cost-effective TLDs to combat phishing campaigns.
% Our result shows that domains targeting certain brands (\ie, \cc{USPS} and \cc{OZON}) use lower-cost TLD (\ie, \cc{top} and \cc{tk}) to register their domains instead using their original benign TLD (\ie, \cc{.com}, \cc{.ru}).
% \KL{highlight if it is a new contribution (surprising), compare with previous work}
% \DK{here.}
}

\PP{Maliciously-registered Vs. Compromised}
The comparison between maliciously registered and compromised domains reveals notable differences in their TLD preferences. 
Among a total of 218,267 \cc{.com} domains, 142,479 (65.3\%) were maliciously registered, while 75,788 (34.7\%) were compromised, indicating that attackers leveraging \cc{.com} domains often register them intentionally for malicious purposes. 
Conversely, new gTLDs such as \cc{.top} and \cc{.xyz} also show a strong preference for malicious registrations, with 54,014 (63.8\%) and 23,085 (61.2\%) domains, respectively, highlighting attackers’ exploitation of low-cost TLDs for scalability. 
In contrast, Freenom TLDs like \cc{.tk} saw relatively balanced usage between maliciously registered and compromised domains before policy changes restricted their availability. 
These patterns suggest that maliciously registered domains favor low-cost or lenient TLDs, while compromised domains may be distributed across a broader range of TLDs, reflecting their opportunistic use of existing infrastructures. 
This distinction underscores the importance of targeted monitoring and stricter enforcement in TLDs that are disproportionately used for malicious registrations.

We analyze the targeted brands between maliciously registered domains and compromised domains.
\cc{Facebook} is the most used targeted brand, with 66,700 phishing domains, of which 58.20\% are maliciously registered. 
\cc{USPS} and \cc{Microsoft} follow, with 41,691 and 26,717 domains, respectively. 
\cc{USPS} exhibits an exceptionally high proportion of maliciously registered domains (90.03\%), indicating that attackers targeting this brand prefer creating new domains rather than compromising existing ones. 
Microsoft demonstrates a more balanced split, with 51.21\% malicious registrations and 46.55\% compromised domains, suggesting a dual approach in leveraging both new and existing infrastructures.

\rtbox{
\textbf{Takeaway:} 
New gTLDs (\eg, \cc{.top}, \cc{.xyz}) are more prevalent in maliciously registered domains, while compromised domains favor legacy gTLDs (\eg, \cc{.net}, \cc{.info}). Freenom TLDs like \cc{.tk} and \cc{.ml} have declined, while \cc{.cn} has increased in 2024.
}

\begin{figure*}[!t]
\centering
    \begin{subfigure}{0.49\textwidth}
        \includegraphics[width=\linewidth]{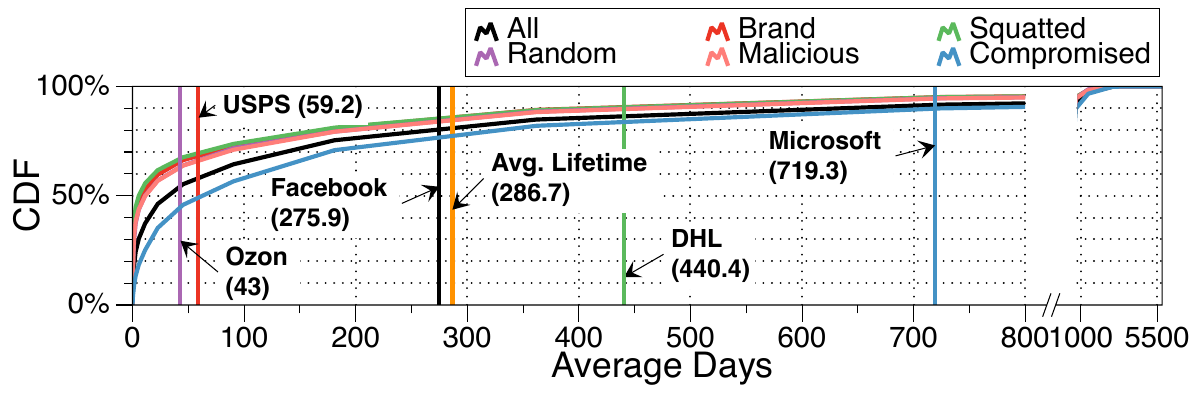}
        \vspace{-20px}
        \caption{Average Delays (days) Between Registration and Detection.}
        \label{fig:registration_detection_avg}
    \end{subfigure}
    \begin{subfigure}{0.49\textwidth}
        \includegraphics[width=\linewidth]{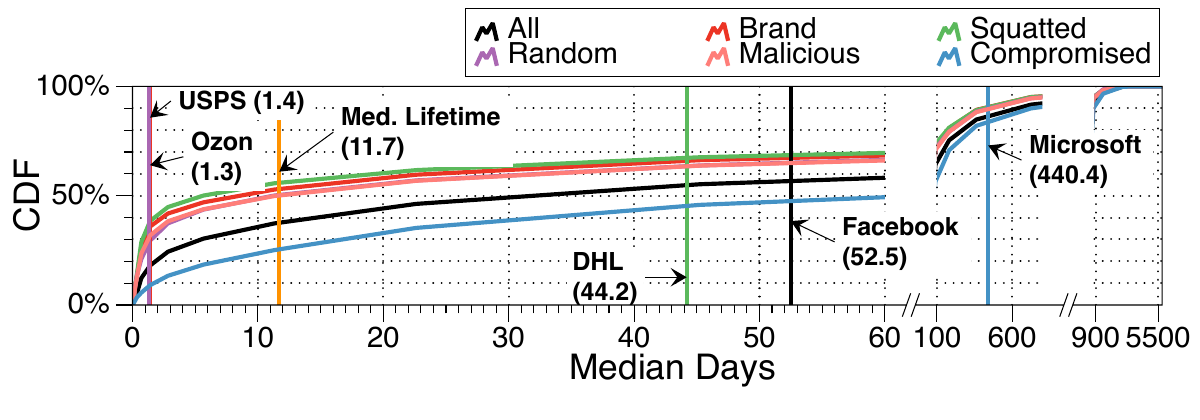}
        \vspace{-20px}
        \caption{Median Delays (days) Between Registration and Detection.}
        \label{fig:registration_detection_median}
    \end{subfigure}
    \vspace{-5px}
    \caption{Delays (days) Between Registration and Detection. Vertical bars show average (or median) days between registration time and detection time of the top 5 most targeted brands.}
    \label{fig:registration_timediff}
    % \vspace{-10px}
\end{figure*}

\subsection{DNS Records}
% \KL{Most used CNAME, NS, MX, TXT}
% \KL{Compare between All vs Mal}
% \KL{Change of A record over time}
% \KL{TTL less than 60 seconds? --> fast-flux dns}
%To understand the characteristics of maliciously registered domains, we look at DNS records of maliciously registered domains. 
We characterize the DNS records of maliciously registered domains collected by our DNS crawler.

% \KH{Is there any comparison between malicious vs. benign domain DNS patterns??}\KL{Just in malicioius}
\looseness=-1

\PP{DNS Records}
We study the values of
commonly used DNS record types
%frequently used values for each DNS record type
(\eg, \cc{A}, \cc{AAAA}, \cc{CNAME}, \cc{NS}, \cc{MX}, and \cc{TXT}). 
Phishing attackers often configure DNS records to evade detection, frequently altering them using techniques such as fast-flux DNS. 
Our analysis reveals that 21.4\% of domains exhibit record changes, with an average frequency of 79.4 days and a median of 125.2 days.
% \KH{States statistics but doesn't connect them to evasion techniques or compare against normal domain behavior patterns.}\KH{DNS record changes percentage appears in multiple places without clear connection to previous findings (64.3\% (here), then 63\% (contributions), then 86.0\% (TTL section)).}

%To gain deeper insights into DNS record changes, we analyze the DNS records of the collected phishing domains. 
We study the types of DNS records configured in phishing domains.
\cc{NS} records were the most common, with a total of 51,459, followed by \cc{A} records (13,218), \cc{SOA} records (8,960), and \cc{TXT} records (5,573).
Focusing specifically on maliciously registered domains, we specifically examined those that exhibited DNS record changes. 
Our analysis shows that only 4.6\% of these domains (117 out of 2,550) demonstrated record changes over time. 
This suggests that modifying DNS records is not a commonly used tactic among maliciously registered phishing domains.
\looseness=-1

To understand the scenarios behind DNS record changes, we manually reviewed domains that exhibited such changes over time. 
One common case involved NS record changes, where domains shifted from one DNS provider to another (\eg, from Cloudflare to Google). 
%Our analysis suggests that these 
Such changes are often motivated by the desire to leverage specific services offered by different DNS providers. 
For instance, attackers may switch to providers like Cloudflare to utilize features, such as free SSL certificates, which are available for a limited duration~\cite{Cloudfla20:online}.

Our analysis reveals that phishing domains show a strong preference for hosting on Amazon Web Services (AWS) infrastructure. 
Specifically, we extract all IP addresses associated with \cc{A} records and utilize the \cc{Summarize IP} feature provided by IPinfo~\cite{IPSummar22:online} to gain insights into their hosting characteristics. 
Among the Autonomous System Numbers (ASNs) analyzed, AS16509 (Amazon.com, Inc.), a primary ASN for AWS, hosts 81.2\% of the phishing domains, while an additional 15.4\% are hosted on AS14618 (Amazon.com, Inc.), another AWS-associated ASN. 
Combined, these two ASNs account for 96.6\% of all analyzed phishing domains, indicating a significant reliance on AWS services. 
This preference may be attributed to AWS’s scalability, cost-effectiveness, and global reach, which make it an attractive option for attackers to host phishing domains. 
In comparison, other hosting providers, such as Google LLC (1.5\%), JSC Selectel (0.2\%), and DigitalOcean, LLC (0.1\%), host far fewer phishing domains. 

\PP{Vantage Point of DNS Server}
%\KL{check}
Phishing attackers can configure location-aware DNS responses.
%to vary based on the victim's location. 
This allows attackers to deliver localized phishing content (\eg, Spanish-language phishing pages for victims in South America) or to evade detection by serving benign pages when accessed from certain locations commonly used by detection systems.

Our preliminary analysis shows that some phishing domains adapt their content to different languages based on the location of the user accessing them. 
However, we do not find any evidence that these phishing domains alter their DNS records based on the vantage point of the queried DNS servers. 
Instead, further investigation reveals that these domains implement language customization through client-side code rather than DNS configuration.

\PP{TTL in DNS Records}
In DNS records, the time-to-live (TTL) specifies how long DNS settings are cached before they are automatically refreshed. Typical TTL values are 12 or 24 hours, with recommended minimum and maximum values of 1 hour (3600 seconds) and 24 hours (86400 seconds), respectively~\cite{DNSTTLbe81:online}. 
%Our analysis reveals that 
% 86.0\% of phishing domains change their TTL values. \RM{change from what to what?}
%, while 14.0\% retain the same value. Additionally, 
2.1\% of the domains use TTL values less than 60 seconds, and 25.8\% use values shorter than 1 hour (3600 seconds) from our dataset. 
Only 2.9\% of the domains set TTLs longer than 12 hours, and among those, 31 domains set values between 12 and 24 hours. 
The median TTL value across domains is 3,994 seconds, while the average is significantly higher at 60,827 seconds. The use of short-lived TTLs can facilitate fast-flux DNS, a technique that frequently changes IP addresses to evade detection~\cite{bilge2011exposure,galloway2024practical} and often employ by
attackers ~\cite{dagon2008corrupted}.
%This behavior, commonly associated with malicious activity, has also been observed in previous research~\cite{dagon2008corrupted}, further highlighting its role in phishing campaigns.
\looseness=-1

\rtbox{
\textbf{Takeaway:} 
Our analysis finds that 21.4\% of domains change their DNS records frequently.
2.1\% of phishing domains configure their DNS TTL values to less than 60 seconds, a configuration commonly associated with fast-flux DNS techniques.
% Our results reveal that \hl{XX} domains exhibit changes in their DNS \hl{XX} domains that are configured to provide different DNS records depending on the location of the request.
}

\section{Lifespan of Phishing Domains}

\label{sec:lifespan}

% For phishing attackers to maximize their monetization, keeping their phishing websites active for as long as possible is essential. To understand this, we analyze how phishing attackers register their domains and examine the timelines from registration to detection by blocklists (\eg, APWG) and eventual deregistration.
This section examines the lifecycle of phishing domains, focusing on two critical phases: (1) the time from registration to detection, (2) the time from detection to deactivation, and (3) the comparison of detection time between blocklists. 
These phases provide insights into how phishing attackers sustain their domains to maximize monetization and evade timely countermeasures. 
By analyzing detection delays and post-detection persistence across different domain types, brands, and registration strategies, we uncover characteristics in the lifespan of phishing domains (\ie, maliciously registered). 
Our findings highlight the need for improved detection mechanisms to reduce delays and more robust enforcement measures to ensure rapid domain takedown, thereby limiting attackers' ability to exploit these domains.

\begin{figure*}[!t]
    \centering
    \includegraphics[width=.98\linewidth]{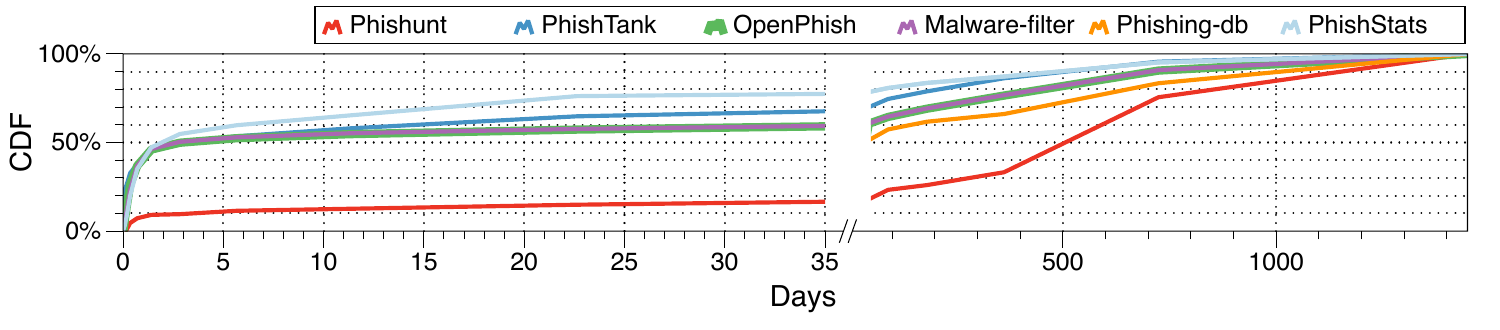}
    \vspace{-15px}
    \caption{Days Between APWG Detection and Other Blocklists.
    Other than Phishunt, all 5 blocklists show similar median delays (2.3 to 4.4 days except the Phishunt).
    }
    \label{fig:blocklist_compare}
    \vspace{-10px}
\end{figure*}

\subsection{Time Taken between Registration to Detection (Detection Delay)}
\label{sec:registration_detection}
In this section, we analyze how phishing domains are detected by blocklist after registration.

\PP{Motivation}
Maliciously registered domains can be blocked in advance when compared to compromised domains.
Phishing domains exhibit significant variation in the time it takes to be detected after registration, influenced by the type of domain and the targeted brand. As shown in ~\autoref{fig:registration_timediff}, these differences highlight both quicker detection for some brands and prolonged delays for others. 

\PP{Result: Overview of Detection Delay}
Across all domains, the overall median detection time is 42.4 days, with an average of 286.2 days.
For the top 10 most targeted brands, the detection times have a slight improvement over these values, with an average of 286.2 days and a significantly shorter median of 11.7 days. This suggests that well-known brands tend to benefit from quicker median detection times compared to less prominent brands, likely due to more active monitoring and stronger anti-phishing measures.

\PP{Detection Time between Targeted Brands}
\cc{USPS} (United States Postal Service)~\cite{WelcomeU93:online}, a U.S. federal agency providing postal services, and \cc{OZON}~\cite{OZON}, a Russian e-commerce platform founded in 1998, stand out with the fastest average detection times among targeted brands. 
\cc{USPS} has a median of 1.4 days (average of 59 days), and \cc{OZON} has a median of 1.3 days (average of 42.9 days).
These quicker detections may result from more active monitoring systems or simpler phishing tactics that are easier to identify. 

Both brands are targeted using non-original TLDs (\autoref{sec:tld_characteristics}), which are often cheaper to register. 
Also, the detection as shown in~\autoref{fig:registration_by_brand}, detection time of \cc{USPS} and \cc{OZON} is quickest with a median of 1.4 days and 1.3 days respectively.
Some registrars, such as Freenom, provide APIs for the immediate takedown of phishing domains upon detecting signs of abuse ~\cite{affinito2022domain}. 
This suggests that attackers' choice of cost-effective TLDs may have inadvertently backfired, as these domains could be removed quickly.
%are detected more quickly than others.
% USPS and OZON are detected the fastest, with average detection times of 59 and 42.9 days, respectively. 
% Interestingly both brands are using non-original TLD. 
% Previous research~\cite{affinito2022domain} highlights that some registrars, such as Freenom, offer APIs to facilitate the immediate takedown of phishing domains upon detecting signs of abuse.

Domains targeting \cc{Microsoft} take the longest to be detected, with an average detection time of 719 days (median of 440.4 days).
\cc{Facebook}, despite being the most impersonated brand, has a moderate detection time of 275 days (median of 52.5 days).
% \DK{put the general median values.} \KL{added in the beginning of this subsection}
% \DK{No you didn't add all median values for the numbers.}
% Microsoft experiences the longest detection delays, with an average of 719 days, reflecting either more sophisticated evasion tactics or less stringent monitoring for its targeted domains. 
% Although Facebook is the most frequently impersonated brand, it does not benefit from rapid detection either, with an average detection time of 275 days. 
These findings highlight significant disparities in detection efficiency across brands and TLDs, emphasizing the impact of attackers' TLD choices on detection timelines.\looseness=-1
% \vspace{-2px}
% This shows that attackers may have chosen to use a cheaper TLD to register a domain, however that backfired on their domain since those domains gets detected sooner than others.
% In contrast, Microsoft experiences the longest detection delay, averaging 719 days. 
% Although Facebook is the most frequently impersonated brand, it does not benefit from quick detection, with an average detection time of 275 days.

% We utilize the Levenshtein distance to find similar domain names.
% From previous work~\cite{maroofi2020comar}, having a Levenshtein distance less than 1 can be considered a similar domain.
% We find that 7.9\% (54,787/689,492) of domains have Levenshtein distance of 1 or less.
% From there, we also look at the registration timestamp and registrars.
% We find that 20.1\% (11,016/54,787) domains are registered within a same time (same time between more than two domains) and also registered through the same registrars.

\PP{Maliciously-registered Vs. Compromised}
As shown in~\autoref{fig:registration_timediff}, the detection times vary across different categories of maliciously registered domains. 
Overall, compromised domains consistently have slower detection times compared to maliciously registered domains, though the difference is not substantial. 
Specifically, the median detection time for maliciously registered domains is 16.3 days, with an average of 206.4 days, while compromised domains have a median detection time of 86 days and an average of 332.1 days. 
This indicates that current detection methods do not perform significantly better at identifying maliciously registered domains compared to compromised domains. We will discuss potential future directions in~\autoref{sec:discussion}.
% mal median: 16.3
% mal avg: 206.4
% comp median: 86
% comp avg: 332.1

% \begin{figure*}[!t]
%     \centering
%     \includegraphics[width=.98\linewidth]{fig/apwg_vs_others.pdf}
%     \vspace{-15px}
%     \caption{Days Between APWG Detection and Other Blocklists.
%     Other than Phishunt, all 5 blocklists show similar median delays (2.3 to 4.4 days except the Phishunt).
%     }
%     \label{fig:blocklist_compare}
%     \vspace{-10px}
% \end{figure*}

\rtbox{
\textbf{Takeaway:} 
Detection delays for phishing domains vary, with maliciously registered domains detected faster (median 16.3 days) than compromised ones (median 86 days). 
Brands like \cc{USPS} and \cc{OZON} see rapid detection (medians of 1.4 and 1.3 days), while others, like \cc{Microsoft}, face significant delays (median 440.4 days and 52.5 days, respectively). 
% Phishing domains from our dataset often use TLDs different from the original brand’s TLD, such as \cc{USPS} being targeted primarily with \cc{.top} instead of \cc{.com}, and \cc{OZON} with \cc{.tk} instead of \cc{.ru}. 
% Detection times also vary significantly by targeted brand; \cc{USPS} and \cc{OZON} are detected the fastest, while Microsoft experiences the longest detection delays.
}

\begin{figure*}[!t]
    \centering
    \includegraphics[width=.98\linewidth]{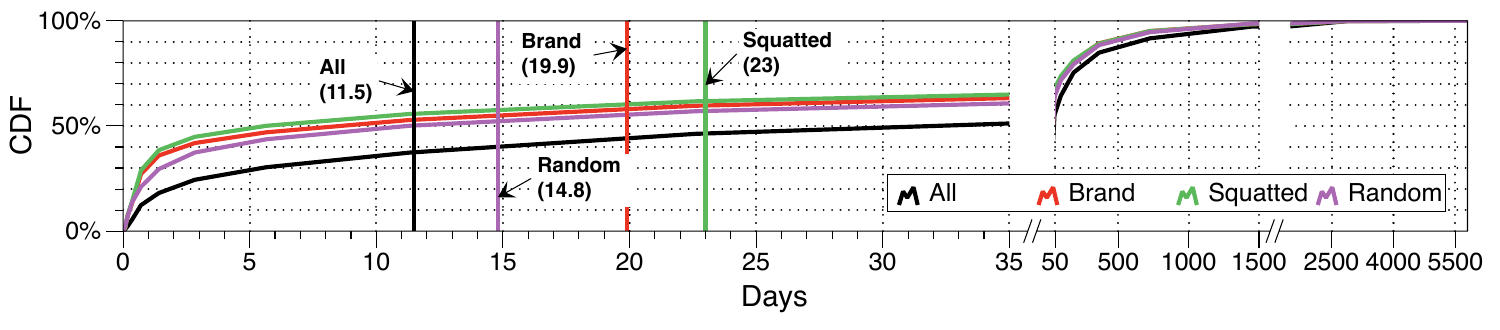}
    \vspace{-15px}
    \caption{Days Between Detection and Last Seen. Vertical bars show the last seen timestamp from the zone file by each type of maliciously registered domain.}
    \label{fig:registration_timediff_last_seen}
    \vspace{-10px}
\end{figure*}

% \subsection{Takendown Domains}
\subsection{Time Taken between Registration and Deregistration (Takedown Delay)}
\label{sec:registration_deregistration}
% \DK{between registration and deregstration? or between detection and deregistration?}
% \KL{Add Deregistration result}
% If domain is still alive after detection, it can still be accessed from victims. Once domain is determined as a malicious, it is important to deregister as soon as possible. 
% \KL{Previous work defined maliciously-registered domain based on lifetimem of phishing domain such as less than 30 days or 60 days. However our finding show that maliciously-registered domain can live up to years.}
\autoref{fig:registration_timediff_last_seen} highlights the significant variation in the time it takes for phishing domains to be deregistered after detection. 
Across all phishing domains, the average time between detection and deregistration is 11.5 days on average, reflecting a relatively short-lived post-detection activity. 
However, specific domain categories reveal notable discrepancies. 
Squatted domains
%, which mimic well-known brands through minor alterations, 
persist significantly longer, with an average lifespan of 23 days post-detection.
Random-looking domains and impersonating specific branded domains exhibit average post-detection duration of 14.8 days and 19.9 days, respectively.
\looseness=-1
% indicating that attackers use different tactics to maintain availability.\KH{This looks slightly overclaimed.}

This prolonged availability of squatted and brand-targeted domains underscores their continued risk in phishing campaigns, as these domains remain accessible to victims even after being blocklisted. The observed differences in deregistration times between maliciously registered domain categories, such as brand-targeted (19.9 days), random-looking (14.8 days), and squatted domains (23 days), may reflect variations in the policies or practices of registrars and hosting providers. 
These differences could also indicate that attackers exploit specific domain types for their perceived resilience or due to differences in enforcement or takedown mechanisms. 
% Further investigation is needed to understand the factors influencing these variations in deregistration times.
These findings reveal critical gaps in enforcement mechanisms, particularly for squatted domains, which outlast other categories by a wide margin. 
% The persistence of these domains highlights the need for more robust takedown processes that prioritize high-risk categories.
% such as squatted domains.
% Strengthening cooperation between registrars, hosting providers, and blocklists could significantly reduce the post-detection window, limiting attackers' ability to exploit these domains further.
% Even after detection by blocklists, phishing domains often remain active for a significant period, allowing attackers to continue exploiting victims. 
% ~\autoref{fig:registration_timediff_last_seen} shows that squatted domains persist, on average, for 23 days post-detection, compared to 11.5 days for all phishing domains. 
% This extended post-detection activity reflects attackers’ ability to prolong domain availability, particularly when targeting specific brands. 
% For instance, domains impersonating Facebook or Microsoft tend to have longer lifespans post-detection.

% The persistence of phishing domains after detection highlights inefficiencies in current enforcement mechanisms. 
% Strengthening coordination between blocklists, registrars, and hosting providers is crucial to ensure the timely takedown of flagged domains.

\rtbox{
\textbf{Takeaway:} 
Maliciously registered domains, especially squatted domains, are key in phishing domains but are deregistered more slowly, averaging 23 days compared to 11.5 days for all phishing domains.
}
\vspace{-5px}

\subsection{Comparison Between Blocklists}
As shown in~\autoref{fig:blocklist_compare}, detection times vary significantly between APWG and other blocklists, illustrating how quickly each blocklist identifies phishing domains after they have already been detected by APWG. 
APWG plays a critical role in identifying phishing domains, with domains on its blocklist having an average detection time of 277.3 days and a median detection time of 42.4 days.

In contrast, other blocklists show considerable delays in detecting these same domains. 
For instance, \cc{Phishunt.io} has an average detection delay of 676.1 days and a median of 930.8 days after APWG's detection, indicating significant lag. 
Conversely, blocklists like \cc{PhishTank} and \cc{OpenPhish} demonstrate faster detection times, with \cc{PhishTank} averaging 167.7 days and a median of 4.4 days, while \cc{OpenPhish} averages 257.9 days and a median of 4.1 days after APWG detection. 
\cc{Malware-filter} and \cc{PhishStats} also detect domains relatively quickly, with median delays of 3.7 and 2.3 days, respectively, despite higher average delays of 255.6 and 141.9 days.
\looseness=-1

\cc{Phishing.Database} shows mixed results, with an average detection delay of 388.5 days but a stronger median delay of 64.4 days. 
These findings demonstrate that APWG consistently detects phishing domains earlier than all other blocklists in our dataset. 
However, the variability in detection delays across blocklists highlights the need for improved synchronization and data sharing to reduce detection gaps and enhance phishing defense coverage. APWG’s early detection could be further leveraged to accelerate response times across the ecosystem.

\rtbox{
\textbf{Takeaway:} 
APWG consistently detects phishing domains earlier than other blocklists, but significant variability in detection delays across blocklists underscores the need for improved synchronization and data sharing to enhance timely phishing defense and reduce attacker impact.
}
\vspace{-5px}
% apwg vs others
% apwg avg: 277.3
% apwg median: 42.4
% phishunt avg: 676.1
% phishunt median: 930.8
% phishtank avg: 167.7
% phishtank median: 4.4
% openphish avg: 257.9
% openphish median: 4.1
% phishtank avg: 167.7
% phishtank median: 4.43
% malware-filter avg: 255.6
% malware-filter median: 3.7
% phishing-db avg: 388.5
% phishing-db median: 64.4
% phishstats avg: 141.9
% phishstats median: 2.3

% \section{Evade Detection in Phishing Domains}
% \label{sec:measurement3}
% We aim to see if phishing domains evade detection by modifying DNS records.
% We first look at any changes in DNS records.
% Then we compare the detection point with the DNS record change to determine whether phishing domains change the DNS record to evade detection.

% \KL{Add trend graph with detection point}

% We also want to see if phishing attackers set different DNS records based on different vantage points around the globe.
% We leverage different DNS servers to assess any inconsistent DNS records between same phishing domain.
% \KL{Add ipinfo map figure}
\section{Discussion}
\label{sec:discussion}
Based on our analysis, we outline limitations and provide recommendations to guide future research efforts.

\PP{Limitation}
During our verification step, a small number of domains may fall outside our defined malicious domain classification categories. 
While we conducted manual verification to ensure the accuracy of our results, it is still possible that a few domains can exhibit characteristics that do not align with our predefined criteria. 

\PP{Recommendation}
There have been various approaches to understanding how phishing attackers exploit domain registration systems and policies to register malicious domains. Previous research has proposed multiple strategies to address this issue, but the persistence of maliciously registered domains indicates that existing efforts remain insufficient. Several approaches have been discussed in prior work to prevent attackers from registering malicious domains:
\begin{itemize}[leftmargin=*, topsep=0pt, itemsep=0em]
    \item \textbf{Stricter Verification Processes}: Implementing enhanced registrant verification during domain registration, such as requiring government-issued identification or multi-factor authentication, to ensure the legitimacy of registrants.
    \item \textbf{Monitoring and Reporting Systems}: Developing real-time monitoring tools to detect suspicious registration patterns, such as bulk registrations or domains containing high-risk keywords, and establishing automated reporting mechanisms to notify registrars and relevant authorities.
    \item \textbf{Registrar Accountability}: Encouraging or mandating registrars to adopt anti-abuse policies, \eg,  proactive detection measures and swift suspension of flagged domains.\looseness=-1
    \item \textbf{Global Collaboration}: Promoting coordinated efforts between registries, registrars, security organizations, and governments to standardize policies and share intelligence on malicious registration practices.
    \item \textbf{Policy Enforcement for Low-Cost TLDs}: Strengthening oversight for TLDs with low registration costs, which are often exploited by attackers.
    % due to lenient policies. \RM{low cost= lenient?}\looseness=-1
\end{itemize}
% 1) Stricter Verification Processes: Implementing enhanced registrant verification during domain registration, such as requiring government-issued identification or multi-factor authentication, to ensure the legitimacy of registrants.
% 2) Monitoring and Reporting Systems: Developing real-time monitoring tools to detect suspicious registration patterns, such as bulk registrations or domains containing high-risk keywords, and establishing automated reporting mechanisms to notify registrars and relevant authorities.
% 3) Registrar Accountability: Encouraging or mandating registrars to adopt anti-abuse policies, including proactive detection measures and swift suspension of flagged domains.
% 4) Global Collaboration: Promoting coordinated efforts between registries, registrars, security organizations, and governments to standardize policies and share intelligence on malicious registration practices.
% 5) Policy Enforcement for Low-Cost TLDs: Strengthening oversight for TLDs with low registration costs, which are frequently exploited by attackers due to lenient policies.
However, due to the decentralized nature of domain registration systems and varying policies among registries and registrars, it is challenging to implement a generalized defense mechanism. Our analysis aims to reiterate these recommendations and emphasize the urgent need for domain registries and registrars to defend against malicious domains proactively. 
By adopting these measures, stakeholders can significantly reduce phishing attackers' exploitation of domain registration systems.

\PP{Ethics}
% This work does not raise any ethical issues.
Our methods emphasize ethical responsibility while upholding scientific rigor in analyzing real-world phishing domains. 
The data collection process including crawling DNS data and registration data (\eg, RDAP), strictly adheres to established ethical guidelines, utilizing phishing URLs sourced from blocklist feeds explicitly made available for research purposes.
\vspace{-10px}
% \RM{I do not see any critical ethics concern, except your crawling speed. I would just put `This work does not raise any ethical issues' for space.}
% Our research methodology prioritizes ethical responsibilities while maintaining scientific rigor in analyzing real-world phishing domains. We implemented key ethical safeguards: collecting data only from established blocklist feeds explicitly available for research purposes, securely storing and anonymizing all collected data, following responsible disclosure practices with registrars and hosting providers, and designing our crawling methods to minimize the impact on legitimate infrastructure. Our analysis methods avoided direct interaction with active phishing sites to prevent inadvertently supporting criminal operations. We conducted this research in compliance with institutional ethics guidelines and cybersecurity research best practices to benefit the broader security community by improving understanding of phishing infrastructure while protecting potential victims.
\section{Related Work}
The number of reports showed the trend of phishing domains and examined phishing websites. However, the characteristics of DNS settings of phishing domains are not well studied.

% \PP{DNS Measure}
% Previous works focus on measured DNS behavior on benign domains~\cite{}.
% Our work focuses on understanding the DNS behavior of phishing domains
% We also focus on patterns of DNS records.

% \PP{DNS in Phishing}
% Previous works:
% Focus on phishing detection with emphasis on evasion techniques.

% Our work:
% Analyze the DNS level on phishing domains to see how phishing attackers register and get detected by blocklists.

% \PP{Phishing Attack Measure}
% Previous researches study on how phishing websites and detection mechanism performs~\cite{}.
% Especially ~\cite{} 

\PP{Coverage of TLDs in Phishing Domains}
% Previous work~\cite{moura2024characterizing} analyzed impersonated domains but limited their focus to three specific ccTLDs, restricting the generalizability of their findings. 
Previous work~\cite{moura2024characterizing} has a narrow focus on three specific ccTLDs, while our work broadens the analysis to include all TLDs (gTLDs and ccTLDs), offering a more comprehensive understanding of phishing trends across a diverse range of domain spaces. 
While the prior study~\cite{moura2024characterizing} focused on identifying impersonated domains, our work delves into how long these phishing domains remain active post-registration, comparing this lifespan to detection delays on blocklists (\eg, APWG). 
Furthermore, our work enhances the analysis by examining the characteristics of specific targeted brands in phishing domains. 
Our work also goes beyond registration trends by exploring the influence of economic factors, such as whether cheaper TLDs contribute to the trend of phishing domains. 
% This economic and strategic perspective on registration patterns adds a valuable dimension to understanding how attackers maintain phishing sites.

\PP{Lifecycle and Classification of Phishing Domains}
The previous study~\cite{maroofi2020comar} introduced a classification method that distinguishes between compromised and maliciously registered domains using 32 extracted features. 
However, their analysis is restricted to identifying malicious domains, leaving a gap in understanding domain behaviors beyond real-world phishing datasets. 
Our work builds upon COMAR~\cite{maroofi2020comar} by applying its method more broadly to understand maliciously registered domains with real-world phishing datasets. 
% Previous research~\cite{hao2013understanding} observed that spammers often use bulk registrations within the \cc{.com} TLD, but this analysis was limited to five months with \cc{.com} TLD. 
Similarly, while prior work~\cite{hao2013understanding} examined bulk registrations, it was limited to five months of \cc{.com} TLD data.
% Our work expands on prior studies by analyzing \emph{multiple} gTLDs and ccTLDs, examining how phishing domains exploit various TLDs over time. 
% By investigating maliciously registered domains, their TLD usage, and targeted brands, we offer deeper insights into the phishing ecosystem and show the difference in characteristics between maliciously registered and compromised domains. 
Our work builds upon these studies by analyzing \emph{multiple} gTLDs and ccTLDs, providing comprehensive insights into how maliciously registered domains behave differently from compromised domains across various TLDs and targeted brands.
\looseness=-1

\PP{Registration Patterns and Longevity of Phishing Domains}
Previous research~\cite{oest2020sunrise} reveals that 75\% of victims access phishing webpages before they are detected. 
Building on this, our work examines the lifespan of phishing domains, compares the lifespans across different targeted brands, and analyzes various types of maliciously registered domains to uncover patterns and strategies used by attackers.

\PP{DNS Behavior and Phishing Detection Delays}
Previous work~\cite{hao2011monitoring} found that 55\% of malicious domains are first detected in spam campaigns over a day after registration. Building on this, our study provides a more comprehensive analysis of phishing domains' lifecycles, spanning from registration to detection and deregistration. By examining the delay between domain registration and blocklist detection, we offer a clearer understanding of detection timelines for both maliciously registered and compromised domains, highlighting gaps in current detection mechanisms.

\section{Conclusion}
This study provides a comprehensive analysis of phishing domains, focusing on maliciously registered domains, DNS behaviors, and detection timelines. We find that 66.1\% of domains are maliciously registered, with attackers favoring low-cost TLDs like \cc{.top} and \cc{.tk}, and frequently using hosting services like AWS. 
Maliciously registered domains are detected faster than compromised ones (median 16.3 vs. 86 days), yet significant delays persist across blocklists. 
Some domains could take over a year to be listed. 
Our findings highlight the need for improved blocklist synchronization and monitoring of widely abused TLDs to mitigate phishing threats effectively. 
\label{sec:conclusion}

\bibliographystyle{plain}
\bibliography{reference}

\appendix
\newpage

\section{Example of Bulk Registration.}~\label{apx:bulk_registration}

\begin{figure}[h]
    \centering
    \includegraphics[width=1\linewidth]{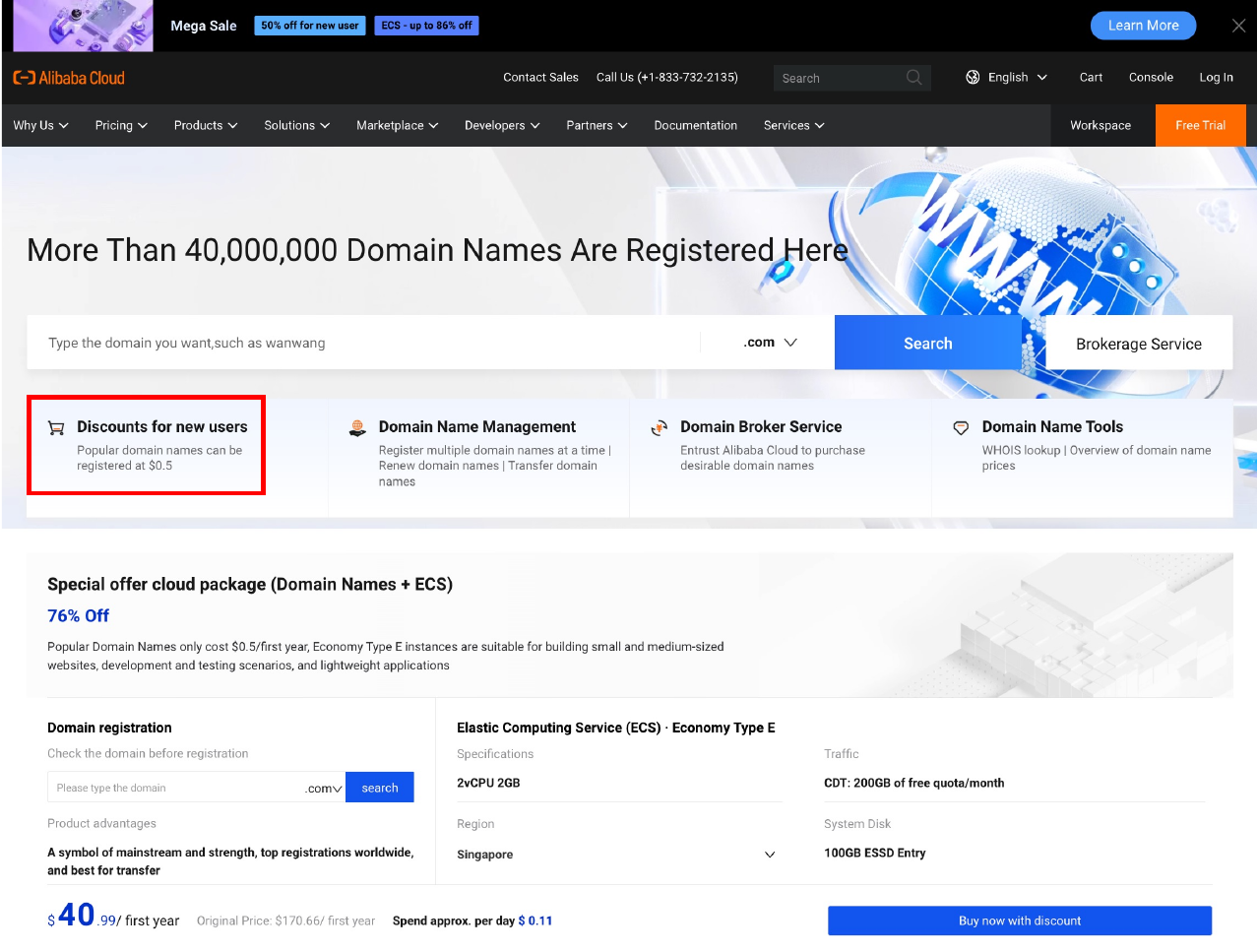}
    \caption{Alibaba Cloud offers promotion and allows bulk registration.}
    \label{fig:alibaba_bulk_registration}
\end{figure}

As shown in~\autoref{fig:alibaba_bulk_registration}, AlibabaCloud provides extensive features that facilitate bulk domain registration while offering promotional pricing for new users. The platform advertises and combines several capabilities that make bulk registration highly accessible: deeply discounted pricing for new users (domains for as low as \$0.5), explicit bulk management tools allowing simultaneous registration of multiple domains, and automation tools for managing multiple domains. While these features serve legitimate business purposes, they can be exploited by attackers to register multiple phishing domains efficiently at minimal cost. The combination of bulk registration capabilities, automation tools, and aggressive pricing promotions makes the platform particularly attractive for malicious actors orchestrating large-scale phishing campaigns.

\end{document}